\documentclass[dvips]{acta}
\usepackage{supertabular,lscape,epsfig}
\usepackage{amssymb}
\usepackage{amsmath}

\SetPages{1}{43}

\SetVol{58}{2008}

\begin{document}

\newcommand{\sa}[1]{\{#1\}}
\newcommand{\sab}[1]{\bigg\{#1\bigg\}}
\newcommand{\ta}[1]{\big<#1\big>}
\newcommand{\tab}[1]{\Big<#1\Big>}
\newcommand{\tabb}[1]{\bigg<#1\bigg>}

\begin{Titlepage}
\Title{ Convective hydrocodes for radial stellar pulsation. Physical and numerical formulation}

\Author{S~m~o~l~e~c, R. and M~o~s~k~a~l~i~k, P.}{Copernicus Astronomical Center, ul.~Bartycka~18, 00-716~Warszawa, Poland\\
e-mail: (smolec,pam)@camk.edu.pl}

\Received{Month Day, Year}
\end{Titlepage}

\Abstract{In this paper we describe our convective hydrocodes for radial stellar pulsation. We adopt the Kuhfu\ss{} (1986) model of convection, reformulated for the use in stellar pulsation hydrocodes. Physical as well as numerical assumptions of the code are described in detail. Described tests show, that our models are numerically robust and reproduce basic observational constraints.

We discuss the effects of different treatment of some quantities in other pulsation hydrocodes. Our most important finding concerns the treatment of the turbulent source function in convectively stable regions. In our code we allow for negative values of source function in convectively stable zones, which reflects negative buoyancy. However, some authors restrict the source term to non-negative values. We show that this assumption leads to very high turbulent energies in convectively stable regions. The effect looks like overshooting, but it is not, because turbulence is generated by pulsations. Also, turbulent elements do not carry kinetic nor thermal energy, into convectively stable layers. The range of this artificial overshooting (as we shall call it) is as large as 6 local pressure scale heights, leading to unphysical internal damping through the eddy-viscous forces, in deep, convectively stable parts of the star.}{hydrodynamics -- convection -- overshooting -- stars: oscillations  -- methods: numerical}

\section{Introduction}

Pulsation hydrocodes play a key role in understanding the variability of classical pulsators: $\delta$~Cephei and RR~Lyrae stars. They enable us to model the light and radial velocity curves and to study the modal selection in these stars.  The first hydrocodes were purely radiative, as it was believed, that convection should not alter the pulsation properties of the stars, specially close to the blue edge. Indeed, radiative hydrocodes were very successful in reproducing many of the observed features (see \eg Buchler 1998 and references therein). However several problems remained (Buchler 1998, Kov\'acs \& Kanbur 1998), with modeling of the double-mode pulsations being the most severe among them. Success of convective hydrocodes in solving this longstanding problem (Koll\'ath \etal 1998, Feuchtinger 1998) focused attention on modeling classical pulsators with convective hydrocodes.

Turbulent convection is an important physical process acting in many types of stars. It is essentially three dimensional process transporting energy through many length-scales: from macroscopic eddy cells to microscopic molecular scales, were energy is dissipated. Stellar convection acts in hard-turbulence regimes (extremely high Rayleigh numbers ($\gtrsim 10^{12}$) and extremely small Prandtl numbers ($\sim 10^{-9}$, Gehmeyr \& Winkler 1992a) which furthermore complicates its modeling. However, many essential features of convection may be described with simple one-dimensional models. Of these, the famous mixing-length theory (MLT) of B\"ohm-Vitense (1958) is most commonly used and underlies many other more complicated models. MLT however, is not suitable for stellar pulsation problems, since it is a local and time-independent theory. In pulsating variable stars, large-scale motions of gas interact with smaller-scale turbulent motions and time-dependent models are necessary to describe the coupling between them. Such one-dimensional models of turbulent convection were developed by many authors. Appropriate equations may be obtained through the Reynolds averaging technique (see \eg Stani\v{s}i\'c 1985). All quantities are decomposed into mean and fluctuating parts, for example for velocity and temperature we have $\boldsymbol{U}=\bar{\boldsymbol{U}}+\boldsymbol{U}'$, $T=\bar{T}+T'$, respectively. Hydrodynamic equations decouple into mean and fluctuating (turbulent) equations, which are coupled by second order correlations, like $\boldsymbol{U}'\boldsymbol{U}'$ or $\overline{\boldsymbol{U}'T'}$, that need to be modeled in order to close the system. Such procedure introduces several dimensionless, order of unity, free parameters, usually denoted by $\alpha$-s. In one-equation models, turbulent equations are reduced to one equation for turbulent energy, $e_t=\overline{\boldsymbol{U}'^2/2}$. An excellent review showing in detail the derivations and approximations made in different models, is given by Baker (1987). We will focus on two models that are commonly  adopted in some recent linear as well as non-linear calculations, namely the Stellingwerf model (1982) and the models based on the work of Kuhfu\ss{} (1986). 

Stellingwerf (1982) truncated the set of three turbulent equations derived by Castor (1968, unpublished) to a one-equation model for the turbulent energy, applying MLT motivated closure relations. To model the small-scale turbulent dissipation, Stellingwerf introduced the eddy-viscous pressure term, in an ad hoc way. Kuhfu\ss{} (1986) model is self-consistent, with all necessary modeling based on physical arguments. Eddy-viscous terms result from first-order modeling of the Reynolds tensor. All the correlations are modeled in a consistent way, using diffusion approximation. This leads to fully differentiable formulation, contrary to the Stellingwerf model, in which effects of buoyant deceleration of the turbulent eddies must be neglected in convectively stable regions. This leads to extreme overshooting in Stellingwerf model as was shown by Gehmeyr \& Winkler (1992a,b). These authors performed a detailed comparison of both models. They favour the Kuhfu\ss{} model pointing other shortcomings of the Stellingwerf treatment. Some drawbacks of the original Stellingwerf theory are overcome by Bono \& Stellingwerf (1992, 1994), who propose a better treatment of convectively stable regions (\cf Section~5). In our hydrocodes we adopt the Kuhfu\ss{} model as it is self-consistent and based on firm physical grounds.

In this paper we present detailed description of our convective hydrocodes. In the first part of the paper (Sections 2 and 3) we discuss the turbulent convection model we adopt in our code: physical formulation in Section 2 and numerical implementation in Section~3 and Appendices. In the second part of the paper (Sections~4~and~5) we present results of some tests of our hydrocode (Section~4), and we discuss the effects of different formulations used in other hydrocodes (Section~5). Conclusions are collected in Section~6.

\section{Physical description of the model}

Our convective hydrocodes implement the time-dependent turbulent convection model proposed, in its original form, by Kuhfu\ss{} (1986). This model was reformulated by Wuchterl \& Feuchtinger (1998) and Feuchtinger (1999) for the use in stellar pulsation hydrocode (Vienna code in the following). Also the hydrocode of Koll\'ath \etal (2002) (Florida-Budapest code in the following) uses the model based on Kuhfu\ss{} derivation (see however Section~5). We use a very similar formulation as in the above mentioned codes, however, we use some different assumptions and somewhat different parametrisation. Therefore, for clarity and completeness, we reproduce below all the equations and quantities we use in our codes. 

Momentum and energy equations are given by
\begin{equation}\frac{d U}{d t}=-\frac{1}{\rho}\frac{\partial}{\partial R}(P+P_t)+U_q-\frac{GM_R}{R^2},\end{equation}
\begin{equation}\frac{d E}{d t}+P\frac{d V}{d t}=-\frac{1}{\rho}\frac{\partial[R^2(F_r+F_c)] }{R^2\partial R}-C, \end{equation}
 \begin{equation}\frac{d e_t}{d t}+P_t\frac{d V}{d t}=-\frac{1}{\rho}\frac{\partial(R^2F_t)}{R^2\partial R}+E_q+C.\end{equation}
Sum of the last two equations form the total energy equation
\begin{equation}\frac{d (E+e_t)}{d t}+(P+P_t)\frac{d V}{d t}=-\frac{1}{\rho}\frac{\partial[R^2(F_r+F_c+F_t)] }{R^2\partial R}+E_q .\end{equation}
In the above equations, $U$ stands for the fluid velocity, which is time derivative of radius, $R$
\begin{equation}U=\frac{d R}{d t}.\end{equation}
$M_R$ is mass enclosed in radius $R$, $V$ is specific volume, which is inverse of density, $\rho$. $E$ and $P$ are pressure and energy of the gas including radiation, while $e_t$ and $P_t$ are turbulent energy and turbulent pressure. Following Wuchterl \& Feuchtinger (1998), we denote viscous energy and momentum transfer rates by $E_q$ and $U_q$ (note however, that Wuchterl \& Feuchtinger 1998 use $U_Q=\rho U_q$ and $E_Q=\rho E_q$). $F_r$, $F_c$ and $F_t$ are radiative, convective and turbulent flux, respectively. The $C$ term describes the coupling between gas energy and turbulent energy equations. Below, we give detailed description of all model quantities.

Turbulent pressure, $P_t$, corresponds to trace part of the Reynolds tensor, while trace-free part leads to turbulent viscosity terms, $U_q$ and $E_q$. These quantities are given by
\begin{equation}P_t=\alpha_p\rho e_t,\end{equation} 
\begin{equation}U_q=\frac{1}{\rho R^3}\frac{\partial}{\partial R}\bigg[\frac{4}{3}\mu_QR^3\bigg(\frac{\partial U}{\partial R}-\frac{U}{R}\bigg)\bigg]=\frac{4\pi}{R}\frac{\partial X}{\partial M},\end{equation}
\begin{equation}E_q=\frac{4}{3}\frac{1}{\rho}\mu_Q\bigg(\frac{\partial U}{\partial R}-\frac{U}{R}\bigg)^2=4\pi X\frac{\partial (U/R)}{\partial M},\end{equation}
where
\begin{equation}X=\frac{16}{3}\pi\mu_QR^6\rho\frac{\partial (U/R)}{\partial M},\end{equation}
and
\begin{equation}\mu_Q=\alpha_m\rho\Lambda e_t^{1/2},\end{equation}
is kinetic turbulent viscosity. $\Lambda=\alpha H_P$ is mixing-length, and $H_P=-\partial R/\partial\ln P$ is the pressure scale height.

The coupling term, $C$, is given by
\begin{equation}
C=S-D-D_{r}.
\end{equation}
The $D$ term describes the dissipation of turbulent kinetic energy into thermal energy (turbulent cascade), while $D_{r}$ describes the radiative cooling of the convective elements (see Wuchterl \& Feuchtinger 1998). Both these terms always damp the turbulent motions. The source function, $S$, is responsible for generation of turbulent energy through buoyant forces. It may drive as well as damp the convective motions. $D$, $D_r$ and $S$ are given by
\begin{equation}D=\alpha_d\frac{e_t^{3/2}}{\Lambda},\end{equation}
\begin{equation}D_{r}=\frac{4\sigma\gamma_r^2}{\alpha^2}\frac{T^3V^2}{c_P\kappa H_P^2}e_t,\end{equation}
\begin{equation}S=\frac{TPQ}{c_PH_P}\Pi,\end{equation}
where $\sigma$ is Stefan-Boltzmann constant, $c_P$ is specific heat at constant pressure, $Q=(\partial V/\partial T)_P$ is thermal expansion coefficient and $\kappa$ is opacity coefficient. $\Pi$ is correlation between entropy and velocity fluctuations, which is given by
\begin{equation}\Pi=\alpha\alpha_s e_t^{1/2}c_PY.\end{equation}
In the above expression, $Y$ is superadiabatic gradient (dimensionless entrophy gradient), given by
\begin{equation}
Y=-\frac{H_P}{c_P}\frac{\partial s}{\partial R}=\nabla-\nabla_\mathrm{a}.
\end{equation}
In convectively unstable layers, $Y>0$, part of the total energy is transported by convective flux, $F_c$
\begin{equation}F_c=\frac{\alpha_c}{\alpha_s}\rho T\Pi=\alpha\alpha_c\rho T c_Pe_t^{1/2}Y.\end{equation}
 The turbulent kinetic energy flux, $F_t$, is approximated by
\begin{equation}F_t=-\alpha_t\rho\Lambda e_t^{1/2}\frac{\partial e_t}{\partial R}.\end{equation}
In the adopted model it is the sole cause of the overshooting of turbulent eddies into convectively stable layers.

As was shown by Wuchterl \& Feuchtinger (1998), diffusion approximation used in modeling the $\Pi$ correlation may be violated in some regions of the star. To fix this problem, these authors introduced the concept of convective enthalpy flux limiter, which we keep in our code as an option. In this case, expression (15) should be replaced by the following formula
\begin{equation}\Pi=\sqrt{\frac{2}{3}}\frac{w}{T}e_t^{1/2}F_L\bigg[\sqrt{\frac{3}{2}}\frac{T}{w}\alpha\alpha_sc_PY\bigg],\end{equation}
where $w=E+P/\rho$ is specific enthalpy, and $F_L$ is the flux limiter function.
 
For the radiative transfer one may use time-dependent, detailed treatment (Vienna code), or simpler models, like diffusion approximation (Florida-Budapest code). The first approach is certainly much more accurate and leads to better physical description of the model's structure, specially in the outer parts. However, in classical pulsator's models, time dependent treatment gives essentially the same results as simple diffusion approximation (Kov\'acs \& Kanbur 1998, Feuchtinger, Buchler \& Koll\'ath 2000). Thus, we adopt the diffusion approximation in our hydrocode, which has the advantage of very low numerical costs. In this approximation radiative flux, $F_r$, is given by
\begin{equation}
F_r=-\frac{4\sigma}{3}4\pi R^2\frac{1}{\kappa}\frac{\partial T^4}{\partial M}.
\end{equation}
Radiation pressure, $P_r$, and radiation energy, $E_r$, are included in $P$ and $E$ together with gas contribution: $P=P_g+P_r$, $E=E_g+E_r$. We have $P_r=aT^4/3$ and $E_r=aT^4/\rho$, where $a$ is radiation constant. Pressure, energy as well as other thermodynamic quantities, are calculated as a function of $T$ and $V$ from the equation-of-state (EOS). We use either simple analytical EOS (Stellingwerf 1982), or detailed EOS tables published by the OPAL group (Rogers \etal 1996). For the opacity coefficient we use the Rosseland mean. By default we use OPAL opacity tables (Iglesias \& Rogers 1996), generated for the solar mixture of Grevesse \& Noels (1993). At low temperatures these are supplemented by Alexander \& Fergusson (1994) opacities (see Pamyatnykh 1999).

Above formulae contain eight order of unity parameters. These are: $\alpha$, $\alpha_m$, $\alpha_p$, $\alpha_t$, $\alpha_s$, $\alpha_c$, $\alpha_d$ and $\gamma_r$. Theory provides no guidance for their values, however, some standard values are in use. Parameters $\alpha_p$ and $\alpha_c$ were introduced by Yecko, Koll\'ath \& Buchler (1998), and are not present in the original Kuhfu\ss{} (1986) derivation, were $\alpha_p \equiv 2/3$ and $\alpha_c\equiv\alpha_s$. Neglecting the radiative losses, local static solution of equations (1)-(3) may be reduced to MLT solution if $\alpha_c=\alpha_s=1/2\sqrt{2/3}$ and $\alpha_d=8/3\sqrt{2/3}$ (Kuhfu\ss{} 1987). With radiative losses included in the model, exact MLT solution cannot be reproduced. We follow Wuchterl \& Feuchtinger (1998) who opt for $\gamma_r=2\sqrt{3}$. We will refer to the above quoted values of $\alpha_p$, $\alpha_s$, $\alpha_c$, $\alpha_d$ and $\gamma_r$ as standard values. We stress however, that they are not based on any firm physical considerations and therefore, should be treated as reference values only. In general, values of the $\alpha$-parameters should be chosen to satisfy the observational constraints. However, results published up to date indicate, that no unique set of convective model parameters reproduces models of both RR~Lyrae and Cepheids in stellar systems of different metallicities. In Table~1, we summarize the $\alpha$ parameters present in our model, and we give their relation to parametrisation used in Florida-Budapest code.

\MakeTable{lccp{3cm}}{12.5cm}{Parameters of the discussed convection model. In the third column we give a standard values, as described in Section~2. Fourth column gives the relation between our parametrization and parametrization used in Florida-Budapest code (barred alphas, Koll\'ath \etal 2002). Only $\alpha_s$ relation is not obvious, which results from complicated parametrisation of source function in Florida-Budapest code. No relation is given for the radiative losses, as this effect is treated differently in both codes (see Section~5.1) . For eddy viscosity treatment see Section~5.2.}
{\hline
 quantity & $\alpha$ & std. value & relation to Florida-Budapest code\\
\hline
mixing-length         & $\alpha$   & -               &$\alpha=\bar{\alpha}$ \\
eddy viscosity        & $\alpha_m$ & -               &$\alpha_m=\bar{\alpha_m}$ \\
turbulent pressure    & $\alpha_p$ & $2/3$           &$\alpha_p=\bar{\alpha_p}$ \\
turbulent source      & $\alpha_s$ & $1/2\sqrt{2/3}$ &$\alpha_s=\bar{\alpha_s}^2\bar{\alpha_d}$ \\
turbulent dissipation & $\alpha_d$ & $8/3\sqrt{2/3}$ &$\alpha_d=\bar{\alpha_d}$ \\
convective flux       & $\alpha_c$ & $1/2\sqrt{2/3}$ &$\alpha_c=\bar{\alpha_c}$ \\
turbulent flux        & $\alpha_t$ & -               &$\alpha_t=\bar{\alpha_t}$ \\
radiative losses      & $\gamma_r$ & $2\sqrt{3}$     &- \\
\hline
}

\section{Numerical representation of the model}

In pulsation hydrocodes, hydrodynamic equations are discretized on a mesh, which is either Lagrangean (fixed mass zones) or adaptive. More sophisticated, adaptive grid is very useful in resolving the narrow features present in classical pulsator's models such as shocks or hydrogen partial ionization regions (PIR). These are not very well resolved by Lagrangean mesh. However, the use of adaptive mesh in purely radiative models just smoothed the light curves, not changing their overall shape (Buchler 1998). In convective models, hydrogen PIR is widened by the convective motions and therefore, is numerically less troublesome. Light curves are smooth already with the Lagrangean mesh. Finally, Feuchtinger \etal (2000) compared the results obtained with Florida-Budapest code (Lagrangean version) with the results obtained with Vienna code (adaptive mesh). Linear results as well as light/velocity curves agree very well. Therefore, instead of increasing numerical costs by additional equation for adaptive mesh,  we decided to keep the simple Lagrangean mesh.

Our codes are based on radiative hydrocodes originally written by Stellingwerf (1975) with some later modifications (Kov\'acs \& Buchler 1988). Inclusion of turbulent convection model however, requires significant changes in numerical methods that are used to solve the hydrodynamic equations. Below we describe the numerical schemes we use in our codes: in static model builder (Section~3.1), in the linear code (Section~3.2) and in the nonlinear direct-time integration hydrocode (Section~3.3). By default, all these codes use the same analytical EOS (Stellingwerf 1982) and opacity procedures, the latter adopted from Warsaw-New Jersey stellar evolution code (Pamyatnykh 1999).

\subsection{Construction of the static model} 

In classical pulsators, inner parts of the star do not participate in the oscillations. We model the outer parts of the star, so-called envelope, only. We neglect rotation and magnetic fields. The model is specified by its mass, $M$, luminosity, $L$, effective temperature, $T_{\mathrm{eff}}$, and chemical composition, $X$ and $Z$. We are not bound by evolutionary tracks in choosing these parameters.  We also need to specify the $\alpha$-parameters entering the convection model we use.

The model is divided into $N$ mass zones. All quantities are defined either at the zones (thermodynamic quantities, $T$, $V$, $P$, $E$) or in between, at so-called interfaces ($R$, $U$, $Y$, fluxes; see Appendix~A). Static model is constructed in two steps. In the first step, we construct an initial model without turbulent pressure and turbulent flux ($\alpha_p=\alpha_t=0$), with turbulent energy and coupling term defined at the interfaces. This initial model is constructed by integration of the static equilibrium equations from the surface inward. The final model, with turbulent pressure and turbulent flux included, and with turbulent energy and coupling term defined at the zones, is constructed through the multivariate Newton-Rhapson iterations. Below we briefly describe these two steps of model construction.

In the first step, we integrate the static, time-independent form of equations (1)-(4) ($U\equiv 0,\ d/dt\equiv 0$) from the surface inward. Initial conditions at the surface are: outer pressure, $P=0$, radius of the outermost interface, $R_N=(L/4\pi\sigma T_{\mathrm{eff}}^4)^{1/2}$ and the temperature of the outermost zone, $T_N^4=fT_{\mathrm{eff}}^4$. By default we use $f=1/2$ resulting from the Eddington approximation and keep $f=3/4$ resulting from the exact solution of the gray atmosphere as an option. We decided to use the same mesh structure as was used in radiative codes (see \eg Kov\'acs \& Buchler 1988). Several outer zones, $N_A$, have equal masses, $DMN$, down to the anchor zone, located at the hydrogen PIR. Temperature of this zone, $T_{\mathrm{A}}$, is fixed, $T_{\mathrm{A}}=11000$K (or $T_{\mathrm{A}}=15000$K, see Section~4). In the inner part of the envelope, below the anchor, masses of the remaining $N-N_A$ zones increase geometrically inward, with the common ratio $h$. Temperature of the innermost zone is $T_{\mathrm{in}}$. Such mesh structure is obtained through adjustment of $DMN$ and $h$ during integrations. First, we integrate from the surface down to the anchor zone and adjust $DMN$ to obtain desired temperature in the anchor zone. Then, we integrate from the anchor zone down to the inner boundary. $h$ is adjusted to obtain the desired inner temperature, $T_{\mathrm{in}}$.

In the second step, results of integration serve as an initial guess for the multivariate Newton-Rhapson iteration of the final model (Appendix A, eqs. (36)-(38)). Turbulent flux and turbulent pressure are turned on (if desired in the computed model) and turbulent energy and coupling term are redefined at the zones. We treat the iterations as converged, if relative corrections to temperatures, radii, and turbulent energies in all zones/interfaces are lower than $10^{-10}$. To preserve the mesh structure, iterations are repeated several times with $DMN$ and $h$ being successively changed. $DMN$ is adjusted to match the desired temperature in the anchor zone. Then, to assure the smooth transition from the upper part (zones of equal mass) to the lower part of the envelope (zone mass increasing geometrically inward), zone mass ratio $h$ must be adjusted. It is done in such a way, that the mass of the envelope below the anchor is not changed during the whole iterative procedure. As a result, the inner temperature of the envelope, $T_{\mathrm{in}}$, changes, but only by several Kelvins.

During the iterations we constrain the outer temperature to $T_N^4=fL/(4\pi\sigma R_N^2)$ and allow for free adjustment of the outer radius, $R_N$. This is fully compatible with the outer boundary condition for luminosity in the nonlinear code (\cf Section~3.3). Outer radius increases if turbulent pressure is turned on in the computed model.

For some models with turbulent pressure included or with strong turbulent flux ($\alpha_t>0.1$) we encounter convergence difficulties. These are overcome by gradually increasing the $\alpha_p$ and/or $\alpha_t$ to the desired value during the iterations.

In Appendix~A, we give the numerical representation of equations solved through the Newton-Rhapson iterations and representation of all quantities that enter the static model computation. We also give representation for turbulent viscosity terms, that are not present in the static model, but enter the linear stability analysis and are present in the nonlinear calculations.

\subsection{Linear stability analysis}

The original Stellingwerf (1975) codes implement the Castor (1971) numerical method in the linear analysis. As implementation of convection model into this scheme is very cumbersome, we drop this scheme and solve the full eigenvalue problem. 

We consider equations (1)-(3) and (5), however with energy equation (2) rewritten in the following form
\begin{equation}c_V\frac{dT}{dt}=-\bigg(P+\Big(\frac{dE}{dV}\Big)_T\bigg)\frac{dV}{dt}-\frac{\partial(L_r+L_c)}{\partial M}-C,\end{equation}
where $c_V$ is specific heat at constant volume. The static, not perturbed model is constructed by just described model builder\footnote{However, we filter out the lowest turbulent energies as they generate numerical havoc, manifesting \eg by erratic oscillations of the linear work integral in the inner envelope. We set $e_t=e_0$ if $e_t<e_0$, with $e_0=1$erg/g. We checked that results are independent of $e_0$ value at least in the range $e_0\in(10^{-4},10^6)$erg/g. With $e_0$ below $10^{-4}$erg/g  numerical havoc appears. On the other side, setting the cutoff above $10^6$erg/g we interfere with significant turbulent energies.}. We linearize this system, treating $R$, $U$, $T$ and $e_t$ as basic variables. We assume $\sim\exp(\sigma t)$ dependence of the perturbed quantities, where $\sigma$ is the complex eigenvalue. Resulting eigenvalue problem is solved using canned eigenvalue solver as suggested by Glasner \& Buchler (1993). We define the linear growth rates of the modes as the fractional growth of the kinetic energy per pulsation period: $\eta=4\pi\Re(\sigma)/\omega$, where $\omega=\Im(\sigma)$ is the mode frequency.

Additional equation for turbulent energy generates a new branch of eigenmodes. These are extremely damped with typical growth rates $\eta<-1$. Therefore, these modes are not expected to cause any troubles in nonlinear calculations (\cf Yecko \etal 1998). 

In addition to the frequencies and growth rates of the modes, linear code calculates the radius, temperature, luminosity and turbulent energy eigenvectors. These are simply returned by the eigenvalue solver. Then, pressure work integrals may be simply calculated. For any pressure term, we have (\eg Castor 1971)
\begin{equation}W_P(M_R)=-\pi\int_{M_0}^{M_R}\Im\{(\delta P)^*(\delta V)\}dM,\end{equation}
where integration is extended over the mass of the envelope. Also eddy viscosity contributes to the work integral. In Appendix~C we derive following formulae for eddy-viscous work
\begin{equation}W_{EV}(M_R)=\pi\int_{M_0}^{M_R}\Im\bigg\{(\delta X)^*\bigg[\frac{\delta V}{R^3}-\frac{3V}{R^3}\frac{\delta R}{R}\bigg]\bigg\}dM.\end{equation}
$\delta X$ is easily calculated from already known eigenvectors (see eq. (9)). It is convenient to normalize the work integrals by the kinetic energy of the mode, $E_K$
\begin{equation}E_K=0.5\int \omega^2|\delta R|^2dM,\end{equation}
and such normalized work integrals are presented in the Figures of this paper.

\subsection{Nonlinear code}

In the nonlinear hydrocode, full set of nonlinear equations is integrated forward in time. The numerical scheme is very similar to the original radiative version (Stellingwerf 1975). Equations (1) and (3)-(5) are written in finite difference form, using $\varpi=e_t^{1/2}$ as the basic variable (Appendix B, eqs. (58)-(61)). In order to preserve the total energy of the envelope, time averages that appear in these equations, must be written in a careful way. We base our scheme (described in Appendix B), on the scheme of Fraley (1968) developed for radiative hydrocodes. 

Multivariate Newton-Rhapson iterations are used to solve the full nonlinear system of difference equations. Temperatures, $T_i$, radii, $R_i$, and turbulent energies, $e_{t,i}$, at time $t$ provide the initial guess, and values at time $t+DT$ are iterated.  We use constant time step, corresponding to roughly 600 steps per pulsation cycle. We treat the iterations as converged, if relative corrections to temperatures, $\delta T_i/T_i$, and $\delta \varpi_i/\varpi_i$ are smaller than $10^{-6}$ in all zones. Typically 3 to 6 iterations (up to $\sim 30$ during the contraction phase) are necessary. For some models convergence difficulties are encountered. If convergence is not achieved in 60 iterations, the current iterations (not all computations) are restarted with halved time step. 

Difference equations are supplemented by the boundary conditions. At the inner boundary we have a rigid core of constant luminosity. Turbulent energy is set equal 0 at the inner boundary as well as in the outermost zone, $e_{t,N}=0$. External pressure is set equal zero, and the outgoing luminosity, $L_N$, is given by $L_N=4\pi R_N^2f^{-1}\sigma T_N^4$.

From numerical point of view, turbulent energy equation requires special attention. All its components depend on $e_t$ in some power. Therefore, $e_t=0$ is always a solution. Once $e_t$ equals 0 in some zone, it will stay equal 0, even if convective instability arises (see \eg Gehmeyr \& Winkler 1992a). Different numerical schemes were developed to overcome this problem (Yecko \etal 1998, Gehmeyr \& Winkler 1992b). We use the scheme very similar to that used by Yecko \etal (1998). Specifically, we add additional non-zero term in the turbulent energy equation, by slight modification of turbulent dissipation term, $D$ (eq.(12))
\begin{equation}D=\alpha_d\frac{e_t^{3/2}-e_0^{3/2}}{\Lambda},\end{equation}
where $e_0$ is a small, constant turbulent energy, which we set to $e_0=10^4$erg/g. In Section~5.3 we describe in detail how this correction works.

Inevitable component of radiative hydrocodes is artificial viscosity. It is necessary to handle shocks developing in the models. It acts as additional pressure, spreading the shock through several mass zones. As an unwanted by-product, artificial viscosity limits the pulsation amplitude, being the main dissipative factor. Amplitudes of the radiative models depend on the parameters of the artificial viscosity, that should be adjusted to match the observational constraints. In convective hydrocodes, eddy viscosity provides physical source of dissipation. In principle, artificial viscosity is not necessary, however we keep it in our code. We use modified Neumann-Richtmyer artificial viscosity (Stellingwerf 1975). It is described by two parameters, $C_Q$ which characterizes the strength of additional pressure, and by cut-off parameter, $\alpha_{\mathrm{cut}}$. Additional pressure turns on, only if relative speed of the consecutive zones exceed the local sound speed by fraction given by parameter $\alpha_{\mathrm{cut}}$. In our models we use $C_Q=4$ and high cut-off parameter, $\alpha_{\mathrm{cut}}=0.1$. With such high cut-off, artificial viscosity does not turn on at all in most of the models. It always plays a subdued role, and is never present in the final limit cycle pulsations. 

Nonlinear hydrocode is supplemented with several data processing tools. For the converged limit cycle pulsations the nonlinear work integrals are calculated (see Appendix~C). Also, bolometric light and velocity curves are computed. During the pulsation cycle, photosphere sweeps through several Lagrangean zones. Therefore, photospheric values are extracted by interpolating to the exact black-body condition, $L=4\pi R^2\sigma T^4$. Colour light curves are obtained through applying bolometric correction at each pulsation phase. We compute bolometric correction using static atmosphere models of Kurucz (2005). 

We stress that the nonlinear hydrocode is fully compatible with the linear one. Exactly the same Lagrangean mesh is used in both codes, as well as EOS and opacity procedures and numerical representation of all quantities. It is extremely important if one interprets nonlinear results in terms of linear ones.

\section{Tests of the code}

Some $\beta$~Cephei models, computed with presented hydrocodes, were already published (Smolec \& Moskalik 2007). In this Section we present limited sample of other test calculations we have done. We stress that these are only test calculations, not intended to model real stars. The goal is to show how our codes work, and to show that resulting models are reliable and numerically robust. We focus our attention on fundamental mode classical Cepheids. We consider models with two sets of convective parameters, A and B, given in Table~2. For $\alpha_s$, $\alpha_c$ and $\alpha_d$ we use standard values (\cf Section~2). Set A represents the simplest convective model without turbulent pressure and turbulent flux, while in set B these effects are turned on. In both sets radiative losses are neglected as well as the turbulent flux limiter. Again we stress, that we did not adjust convective parameters to match the observational constraints. For all the models discussed in this paper, we use Galactic chemical composition, $X=0.7$ and $Z=0.02$, and following mass-luminosity relation: $\log (L/L_\odot)=3.56\log (M/M_\odot)+0.7328$ (Szab\'o \etal 2007).

\MakeTable{ccccccccc}{12.5cm}{Two sets of convective parameters considered in this work. $\alpha_s$, $\alpha_c$, $\alpha_d$ and $\alpha_p$ are given in the units of standard values.}
{\hline
Set & $\alpha$ & $\alpha_m$ & $\alpha_s$ & $\alpha_c$ & $\alpha_d$ & $\alpha_p$ & $\alpha_t$ & $\gamma_r$\\ 
\hline
A & 1.5 & 0.20 & 1.0 & 1.0 & 1.0 & 0.0 & 0.0 & 0.0 \\
B & 1.5 & 0.25 & 1.0 & 1.0 & 1.0 & 1.0 & 0.01 & 0.0 \\
\hline
}

 Model envelopes are divided into 150 mass shells extending down to $2.5\cdot 10^6$K. 40 exterior zones have equal mass down to the anchor zone. Mass of the interior zones increase geometrically inward. For models of set A we set the anchor temperature to 11000K. Such anchor choice is not good for models of set B (and generally for models with turbulent flux) as the growth rates are not smooth along the sequence of models (\eg for models of constant mass and differing temperature). This is clearly numerical effect, resulting from poor resolution, as growth rates for models calculated with denser mesh (300 mass shells) exhibit smooth behaviour. Smoothness may also be obtained by setting the anchor temperature to higher value. With 15000K growth rates are smooth. Therefore, we use this value of anchor temperature for models of set B.

Static models constructed by model builder, are subject to linear stability analysis. This allows to determine the pulsation instability strips (IS) in the Hertzsprung-Russel diagram. For the discussed sets of convective parameters, IS are plotted in the left panel of Fig.~1 (set A) and in the right panel of Fig.~1 (set B). Thick solid lines define the fundamental mode IS, while thick dotted lines enclose first overtone IS. For set A, instability strips are shifted by $\sim$200K toward higher temperatures, and are narrower than for set B. However, first overtone IS extends to slightly higher luminosities for set A. At low luminosities, the widths of the fundamental mode IS are roughly $\sim$800K and $\sim$ 950K for sets A and B, respectively. Fundamental mode IS widens toward higher luminosities for both sets of convective parameters.

Structure of typical static model of set B is depicted in Fig.~2. This model has 4.5\MS{} and lies inside the fundamental mode IS, 300K from its blue edge. Upper panel of Fig.~2 shows the run of superadiabatic gradient, $Y=\nabla-\nabla_\mathrm{a}$, and adiabatic gradient, $\nabla_\mathrm{a}$, versus the zone number. Arrows indicate the minima of $\nabla_\mathrm{a}$ connected with partial ionization regions (PIR). Hydrogen and both helium PIR are clearly resolved. These regions give rise to convective instability, that is $Y>0$. In convectively unstable regions, part of the flux is carried by convection (solid line in the lower panel of Fig.~2). In the discussed model turbulent flux (dotted line in the lower panel of Fig.~2) diffuses the turbulent energies (dashed line in the lower panel of Fig.~2) also into convectively stable zones, where $Y<0$ (overshooting). Displayed curves are smooth and all ionization features are properly resolved, which shows that our mesh is reasonable.

Linear stability of the models may be studied through linear work integrals. For the discussed model, fundamental mode work integrals are plotted in Fig.~3. Total work is plotted with solid line, turbulent pressure work with dotted line and eddy-viscous work with dashed line. Upper panel of Fig.~3 shows the local work, that is work done in individual zones over one pulsation period, while lower panel shows the cumulative work integrals (expression (22)). Work integrals are normalized by kinetic energy of the mode. As a result, total cumulative work at the surface, is equal to the growth rate of the mode. Radiative damping acting in the interior of the model is overcome by the driving through the $\kappa$-mechanism acting in the second helium and hydrogen-helium PIRs. Eddy viscosity has always stabilizing effect, while turbulent pressure work may contribute both to damping and driving. In the discussed model its overall effect is neutral (bottom panel of Fig.~3).

Linear stability analysis provides information whether the model is stable or unstable against small perturbations. In case of instability it tells nothing about the final pulsation state, its amplitude, light/velocity curves, or the final modal selection. This is the domain of nonlinear calculations. In these calculations static model is kicked with the scaled velocity eigenvector of the desired mode and time evolution of the model is followed.

The consistency between linear and nonlinear calculations may be checked, by initializing nonlinear calculations in linear regime, that is with small surface velocity amplitude, say 0.1km/s. Then, the growth rate of the initialized mode may be calculated through the nonlinear work integrals (see Appendix~C), as the average total envelope work, over several initial pulsation cycles. Such determination is not very accurate, as our initialization is never clean, that is in addition to the desired mode, also other modes, specially higher order, strongly damped overtones are present in the initial phase of integration. Nevertheless, such calculated growth rates (for both fundamental and first overtone modes) agree with the linear values, typically within $\pm 3$\%, differences larger than $\pm 5$\% appear only exceptionally.     

Full nonlinear calculations were performed for fundamental mode models, with convective parameters of set B. Integrations were carried over several hundred to few thousand pulsation cycles, till the limit cycle pulsation was reached. In Fig.~4 we show the nonlinear work integrals for the already discussed 4.5\MS{} model. These should be compared with linear work integrals displayed in Fig.~3. Artificial viscosity does not contribute to the nonlinear work integrals at all. It is clearly visible in the lower panel of Fig.~4 that pulsation instability is saturated. Total cumulative work integral at the surface is equal to 0. In comparison to linear work integrals, sharp features are widened, being smeared by the motion of ionization fronts through the Lagrangean zones of the model. The strong damping by the eddy viscosity is clearly visible. In practice it means, that the $\alpha_m$ parameter may be used to control the limit cycle amplitude, similar to artificial viscosity parameters in radiative codes. Higher $\alpha_m$, stronger the eddy-viscous damping and lower the amplitude. 

Fig.~5 shows the radial velocity (left panel) as well as bolometric light curves (right panel) calculated for sequence of fundamental mode models of set B. The Hertzsprung bump progression (\eg Buchler, Moskalik \& Kov\'acs 1990) connected with the 2:1 resonance between the fundamental mode and the damped second overtone, $2\omega_0=\omega_2$, is clearly visible. This resonance is crucial in shaping the light/velocity fundamental mode curves at periods around 10 days, were resonance center is located (Simon \& Schmidt 1976, Kov\'acs \& Buchler 1989). Considering velocity curves and starting from shorter periods (left panel of Fig.~5, bottom), the bump first appears at the descending branch of the velocity curve and then, as period of the fundamental mode grows (and $P_2/P_0$ ratio gets smaller), bump moves toward ascending branch and finally disappears. Our radial velocity curves are smooth, which is not the case for bolometric light curves during part of the expansion phase (Fig.~5, right). For massive models series of wiggles appears on the descending branch. These are due to poor resolution of our Lagrangean mesh. As envelope expands, very thin convection zone, sweeps through several mass shells, moving inward the model.

Hertzsprung bump progression, as well as comparison of model curves with observed data, may be best studied through the Fourier decomposition parameters (Simon \& Davies 1983). Here we focus on radial velocity curves. These are decomposed into Fourier series, and amplitude ratios, $R_{k1}=A_k/A_1$, and Fourier phases, $\phi_{k1}=\phi_k-k\phi_1$, are calculated. In Fig.~6 we show the run of $A_1$, $R_{21}$ and $\phi_{21}$ for computed radial velocity curves of set B, running 300K-600K to the red of the blue edge of the fundamental mode IS. Dots represent the observational data of Moskalik, Gorynya \& Samus (2008, in preparation). Despite the fact that we haven't attempted to adjust the convective parameters to match the observational constraints, the overall agreement is quite good. Concerning $A_1$ and $R_{21}$, we note some problems for shorter periods. Specifically, models do not reproduce the sharp increase of $R_{21}$. Concerning $\phi_{21}$, the model curves are shifted toward shorter periods in comparison to observations. This is connected with the location of the  $2\omega_0=\omega_2$ resonance center. The characteristic run of $\phi_{21}$ is directly connected with this resonance (Buchler, Moskalik \& Kov\'acs 1990). Observationally the resonance center falls around 10 days. In numerical models, resonance location depends mostly on chosen mass-luminosity relation and on convective parameters. These were not adjusted to match the resonance center which explains the shift. In practice it is hard to infer the resonance location from observations only. Resonance location may be determined through the fit of the theoretical run of $\phi_{21}$ versus period to the observational data points (\eg Kienzle \etal 1999).

\section{Discussion of approximations and representations used in other hydrocodes}

In different hydrocodes, adopting essentially the same convection model, some processes or model quantities are modeled or treated in different ways. This concerns the modeling of radiative damping of the convective elements, modeling of the eddy-viscous terms and treatment of the turbulent source function and convective flux in convectively stable regions. 

\subsection{Modeling of radiative damping of the convective elements}

We describe the radiative damping of the convective elements through the radiative cooling term, as proposed by Wuchterl \& Feuchtinger (1998) for the Vienna pulsation code (see Section~2). Different approach is adopted in Florida-Budapest code, where P\'eclet correction factor is used (Buchler \& Koll\'ath 2000). P\'eclet factor multiplies the source term as well as convective flux, and accounts for decrease of convective efficiency in the limit of small P\'eclet numbers. As we have not implemented the P\'eclet correction in our hydrocode, we do not discuss the possible differences in computed models, caused by the two described treatments. 

\subsection{Treatment of eddy viscosity}

In different pulsation hydrocodes, eddy-viscous terms are treated in different ways. We use the form derived by Kuhfu\ss{}, resulting from first order modeling of the Reynolds tensor. The same form is used in the Vienna code (Wuchterl \& Feuchtinger 1998). Many workers use eddy-viscous pressure, introduced in an ad-hoc way by Stellingwerf (1982). The form proposed by Koll\'ath \etal (2002), slightly different from the original Stellingwerf (1982) form, is most commonly used (\eg in Florida-Budapest code, Olivier \& Wood 2005, Keller \& Wood 2006). We will refer to these treatments of eddy viscosity as to Kuhfu\ss{} eddy viscosity and Koll\'ath eddy viscosity. If Koll\'ath eddy viscosity is used, $E_q$ and $U_q$ terms are not present in equations (1)-(4), but additional pressure term of the form
\begin{equation}P_{\nu}=-\frac{4}{3}\alpha_m\rho\Lambda e_t^{1/2}R\frac{\partial}{\partial R}\bigg(\frac{U}{R}\bigg),\end{equation}
should be placed next to turbulent pressure term, yielding following momentum equation
\begin{equation}\frac{d U}{d t}=-\frac{1}{\rho}\frac{\partial}{\partial R}(P+P_t+P_{\nu})-\frac{GM_R}{R^2}.\end{equation}
Simple algebra shows, that in the above equation the $-\frac{3}{\rho R}P_{\nu}$ term is dropped in comparison to momentum equation with Kuhfu\ss{} eddy viscosity (eq.(1)).

To check the effects of using eddy viscosity in the Koll\'ath form, we implemented it in our hydrocode. Numerical representation of eddy-viscous pressure, $P_{\nu}$, is given at the end of Appendix~A. Numerical scheme is much simpler, than in case of Kuhfu\ss{} treatment, as $P_{\nu}$ is just additional pressure term, and the Fraley numerical scheme, the same as in purely radiative case, may be used (see Appendix~B).

In Fig.~1 we compare instability strips for models with convective parameters of sets A and B (Table~2) computed with Kuhfu\ss{} eddy viscosity (thick lines), and with Koll\'ath eddy viscosity (thin lines). In the latter case instability strips are narrower. Comparing individual models we note higher growth rates for models computed with Kuhfu\ss{} eddy viscosity. This effect may be compensated by lowering $\alpha_m$ in models computed with Koll\'ath eddy viscosity (keeping other alphas fixed).

There is no unique and representative way to compare the nonlinear models. We decided to compute the sequences of nonlinear models of set B, lying 300K from the blue edges of the instability strips, computed with Kuhfu\ss{} and with Koll\'ath eddy viscosities. As these linear edges do differ (Fig.~1) also effective temperatures of the computed nonlinear models of the same mass do differ. Fourier decomposition parameters for the velocity curves are plotted in Fig.~7, solid line for Kuhfu\ss{} eddy viscosity, dotted line for Koll\'ath eddy viscosity. In agreement with linear results, amplitudes are lower for models computed with Koll\'ath eddy viscosity. Consequently, also $R_{21}$ is lower for these models. Concerning $\phi_{21}$ both treatments of eddy-viscosity give roughly the same results. Despite the lower amplitudes for models computed with Koll\'ath eddy viscosity, qualitative run of Fourier parameters is the same in both treatments of eddy viscosity.

\subsection{Treatment of source function and convective flux in convectively stable regions}

As already mentioned by Kuhfu\ss{} (1986), one of the shortcomings of the Stellingwerf (1982) theory is that the source function, $S$, cannot damp the turbulent motions, when a given layer becomes convectively stable during pulsation. This is due to the $S\sim\sqrt{Y}$ dependence, resulting from the chosen closing relation of the Stellingwerf's convection model (see Section~1). Similarly, for the convective flux, $F_c\sim\sqrt{Y}$, in this theory. Such formulation leads to several problems, specially the range of overshooting is large and characteristic time scales for the growth and decay of turbulent energies cannot be defined in a reasonable fashion (Gehmeyr \& Winkler 1992b). To overcome these problems, Bono \& Stellingwerf (1992, 1994) modified the original Stellingwerf model and set $S\sim\mathrm{sgn}(Y)\sqrt{|Y|}$ and $F_c\sim\mathrm{sgn}(Y)\sqrt{|Y|}$. Kuhfu\ss{} theory is void of these problems. It offers physically well motivated, differentiable formulation, as we have $S\sim Y$ and $F_c\sim Y$ in this model. Thus, in convectively stable regions, both $S$ and $F_c$ have negative values (see subsection 5.3.3 below). This is the formulation applied in our code, and \eg in Olivier \& Wood (2005). However some of the workers using the Kuhfu\ss{} model cut the source function, and in convectively stable regions, $Y<0$, set $S=0$. The same restriction is applied to convective flux. Symbolically we write for such modified-Kuhfu\ss{} model, $S\sim Y_+$ and $F_c\sim Y_+$. This is done \eg in the Florida-Budapest hydrocode (Koll\'ath \etal 2002). Buchler \& Koll\'ath (2000) show, that if $\alpha$-parameters of the convective model are recalibrated accordingly, Stellingwerf (1982) theory ($S\sim\sqrt{Y}$ and $F_c\sim\sqrt{Y}$) and modified-Kuhfu\ss{} theory ($S\sim Y_+$ and $F_c\sim Y_+$), as well as mixed formulation of Yecko \etal (1998) ($S\sim \sqrt{Y}$ and $F_c\sim Y_+$) give very similar results. Below we argue that their conclusion results from the fact, that in all these models source function is equal 0 in convectively stable zones. If we do not limit the source function to non-negative values, both Kuhfu\ss{} and Stellingwerf theories differ significantly, as studied analytically by Gehmeyr \& Winkler (1992b). With our convective hydrocode modified to use non-negative source function we find further differences. Specially, we show that non-negative source function leads to significant turbulent energies in the inner, convectively stable zones. 

Structure of this Section is as follows. In subsection~5.3.1 we compare the results obtained with our code for two cases discussed above: $S\sim Y$ and $F_c\sim Y$ (our default formulation) and $S\sim Y_+$ and $F_c\sim Y_+$ (Florida-Budapest formulation). In subsection~5.3.2 we show, that in case of $S\sim Y$, treatment of convective flux plays only a minor role. In subsection~5.3.3 we summarize the physical arguments justifying treatment used in our code.

\subsubsection{ $S\sim Y$ and $F_c\sim Y$ versus $S\sim Y_+$ and $F_c\sim Y_+$ }

We focus our attention on the simplest models with convective parameters of set A. We discuss two convection models, namely our standard model with $S\sim Y$ and $F_c\sim Y$, and modified one with  $S\sim Y_+$ and $F_c\sim Y_+$. We will refer to these convection models, as well as to pulsation models computed with them, as to NN and PP, respectively. It turns out that linear results for both NN and PP models are almost identical. Respective edges of the IS in the left panel of Fig.~1 would overlap, as they are shifted by less than 0.5K. Full nonlinear calculations were performed for sequence of models lying 300K to the red of the blue edge of the fundamental mode IS. We stress that respective models from NN and PP sequences have the same masses, luminosities and effective temperatures. Despite the fact that linear results are almost identical, nonlinear results differ significantly. This is shown in Fig.~8, where we plot Fourier decomposition parameters, $A_1$, $R_{21}$ and $\phi_{21}$, for both NN (solid line) and PP (dotted line) models. The most striking difference concerns amplitudes, and consequently the $R_{21}$ ratio. For PP sequence amplitudes are significantly lower, specially at shorter periods. This leads to decrease of $R_{21}$ ratio, not observed for NN models. We trace these differences to very different behaviour of the models in the deep, convectively stable ($Y<0$) interior. To explain these differences we focus our attention on one model of 4.5\MS. Figs.~9 display the spatial profiles of turbulent energy during one pulsation cycle. In the upper panel (Figs.~9a,b) profiles of $e_t$ for NN model are plotted, while in the lower panel (Figs.~9c,d) we plot the profiles for PP model. For each model two viewpoints are used to highlight the internal zones (panels a,c of Fig.~9) and external zones (panels b,d of Fig.~9). Note different scales on vertical, logarithmic axes for NN and PP models. In both models internal zones are convectively stable. $Y$ becomes positive in zones around 70, independently of the pulsation phase, and the model considered (NN/PP). In Figs.~10 we show the corresponding nonlinear work integrals for NN model (upper panel of Fig.~10) and PP model (lower panel of Fig.~10). It is clear from Figs.~9~and~10 that in the PP model very high turbulent energies are present in convectively stable zones below zone 70 (zones 40-70), despite the fact that turbulent flux is not turned on. Such high turbulent energies cause significant eddy-viscous damping in the deep interior, as is well visible in the lower panel of Fig.~10. Consequently, amplitudes are lower for PP model. For NN model, turbulent energies are negligible in the discussed internal zones (below zone 70), and hence the eddy-viscous damping is not present there (upper panel of Fig.~10). 

To explain the reasons for such differences we need to analyze the turbulent energy equation for NN and PP case in detail. We rewrite eq. (3) in the following form
\begin{equation}\frac{de_t}{dt}=\underbrace{S}_{S-\mathrm{term}}\underbrace{-\alpha_d\frac{e_t^{3/2}}{\Lambda}}_{D\mathrm{-term}}\underbrace{+\alpha_d\frac{e_0^{3/2}}{\Lambda}}_{e_0-\mathrm{term}}\underbrace{+E_q}_{E_q-\mathrm{term}}.\end{equation}
Turbulent flux and turbulent pressure terms are dropped (we consider set A of convective parameters), and we make use of eq. (25). The time derivative from the left-hand-side is to be balanced by the source term ($S$-term), turbulent dissipation term ($D$-term), correction term ($e_0$-term), and eddy-viscous energy transfer term ($E_q$-term). $E_q$ and $e_0$ terms are always positive and thus, always drive the turbulent energies. $D$-term is always negative and thus, always contributes to the damping of turbulent energies. $S$-term may drive, as well as damp the turbulent energies in case of NN models, and always drive the turbulent energies in case of PP models. It is also important to notice that $S$ and $E_q$ terms depend on $e_t$ like $\sim e_t^{1/2}$, while $D$-term depends on $e_t$ like $\sim e_t^{3/2}$. We also note that $e_0$-term has slightly different form in Florida-Budapest code, where it is equal to $\alpha_de_0e_t^{1/2}/\Lambda$ (Yecko \etal 1998). However, the following discussion concerning PP models, does not depend on the exact form of $e_0$-term, what we checked numerically (by using the $e_0$-term in the form of Yecko \etal 1998) and what will become clear from the discussion below. 

We focus our attention on the bottom boundary of the convectively unstable region, where $Y$ changes its sign (around zone 70), and on the internal, convectively stable regions. Below we show that the different treatment of $S$-term in NN and PP convection models leads to the observed differences in computed pulsation models.

In NN model, as $Y$ becomes negative, also $S$-term becomes negative and together with $D$-term they damp the turbulent energies. As is visible in Fig.~9a, around zone 70, extremely rapid fall of turbulent energies, by roughly 25 orders of magnitude, happens. This effective damping is mainly due to the $S$-term, and its $\sim e_t^{1/2}$ dependence on $e_t$. $D$-term plays only a minor role in the described turbulent energy fall. The driving effect of $E_q$-term is overcome by the damping effect of the $S$-term. Turbulent energies fall rapidly, and only the $e_0$-term prevents them from falling to zero. This is the only role of this term.  We note in passing, that as $S$, $D$ and $E_q$ terms depend on $e_t$ in some power, $e_t=0$ would be a solution of equation (28). As $e_t$ becomes 0 in some zone, it will stay zero, even if convective instability arises. Therefore, we need a non-zero, small term in turbulent energy equation, that would act as a seed for turbulent energies, as convective instability arises. In convectively stable regions, distribution of turbulent energies depends mainly on the balance between $S$, $e_0$ and $E_q$ terms. $D$-term plays negligible role due to its $\sim e_t^{3/2}$ dependence on $e_t$. As is well visible in Fig.~9a, with our choice of $e_0$ ($e_0=10^4$erg/g), turbulent energies in the deep envelope (below zone 70), stabilize at a very small level, $\sim 10^{-12}-10^{-14}$erg/g. This level depends on $e_0$ value, which should be chosen in order to assure very small, non-zero energies in convectively stable regions. This is the case with our choice. Distribution of turbulent energies at this low level is shaped by physics represented \eg in the $S$-term, hence the small bump of $e_t$ around zone 40, caused by the iron opacity bump. Turbulent energies of order of $10^{-12}-10^{-14}$erg/g are negligible and physically not important. On the other side, they are sufficient to rebuilt the high turbulent energies if convective instability arises. This is clearly visible in Fig.~9b. As convective instability sweeps into external zones, turbulent energies are ignited from negligible level ($10^{-10}$erg/g) to full strength ($10^{13}$erg/g). We stress again that the value of $e_0$ determines the very small level of turbulent energies in convectively stable zones of the model, but as such, it is not responsible for it. Without $e_0$ turbulent energies would fall to zero, which is, as mentioned, unwanted. In any case, $e_0$-term does not fasten this fall of amplitudes as it is always a driving term. Therefore, a rapid fall of turbulent energies as $Y$ becomes negative, is entirely caused by physical terms. Thus, $e_0$-term plays only a numerical role, which we checked carefully by varying its value (in reasonable range).

In the PP model situation is different. When $Y$ becomes negative, source term is set equal to 0. The damping by the $D$-term reduces the turbulent energies. As is visible in Fig.~9c, turbulent energies around zone 70 fall roughly by three orders of magnitude from $10^{13}$erg/g to roughly $10^{10}$erg/g. Then, the damping effect of the $D$-term is balanced by the driving effect of $E_q$-term. Note that $D$-term depends on $e_t$ like $\sim e_t^{3/2}$, while $E_q$  depends on $e_t$ like $\sim e_t^{1/2}$. As $e_t$ falls, the damping strength of $D$-term falls more rapidly, than the driving strength of $E_q$-term. In the absence of damping $S$-term (which would depend on $e_t$ in the same way as $E_q$) balance between $D$-term and $E_q$-term sets the turbulent energies on relatively high level, $10^9-10^{10}$erg/g in the zones 40-70 (Fig.~9c). Then, a slow decline of $e_t$ below zone 40 is observed. This decline reflects the vanishing amplitude of the fundamental mode, as one moves inward the model. Consequently $E_q$-term decreases inward the model and vanish at the inner rigid boundary, where $e_t=e_0$ (eq. (28)). $e_0$-term plays negligible role in the described balance and its value (in reasonable ranges) has no effect on the described turbulent energy distribution. Even more. As we restart the nonlinear calculations for the full amplitude model with $e_0$-term dropped, we find no numerical problems, and no change in model behaviour is observed. In convectively stable zones, 40-70, turbulent energies are reduced by only 3 orders of magnitude compared to the values in the center of convective zone, and they are still high enough to produce significant eddy-viscous damping in the deep interior. This damping is clearly visible in the lower panel of Fig.~10, and is responsible for lower amplitudes of PP models visible in Fig.~8.

It is clear from the presented discussion, that the different treatment of source term in convectively stable regions is responsible for the observed differences between NN and PP models. The crucial point is, that in PP model $S=0$ in convectively stable zones. The form of source term in convectively unstable regions  ($S\sim Y_+$ \vs $S\sim\sqrt{Y}$) is not very important as was checked by Buchler \& Koll\'ath (2000). Nonlinear work integrals presented by Buchler \& Koll\'ath (2000) display the same feature as visible in lower panel of Fig.~10 -- strong eddy-viscous damping in more than 30, convectively stable model zones. If we allow for negative values of source function, strong eddy-viscous damping is not present in convectively stable zones. This result does not depend on the exact form of the source function, either. Both with $S\sim Y$ (our approach, upper panel of Fig.~10) and with $S\sim\mathrm{sgn}(Y)\sqrt{|Y|}$ (Bono \& Stellingwerf 1994) eddy viscous damping is not present in the inner, convectively stable parts of the model (see work integrals \eg in Bono, Marconi \& Stellingwerf 1999).

As we described, high turbulent energies in the PP model, are partly caused by the turbulent energy driving through the $E_q$-term. This may be further confirmed, by calculation of NN and PP models without eddy viscosity ($\alpha_m=0$). In the NN model situation in the convectively stable regions does not change. Turbulent energies are still extremely small. In the PP model, in convectively stable zones, $S$-term equals to 0, and $e_t=e_0$ is now a solution. With $e_t=e_0=10^4$erg/g turbulent energies are 9 orders of magnitude smaller in comparison to turbulent energies in the center of convective zones (to be compared with 3 orders of magnitude reduction in case of $\alpha_m \ne0$). Such energies are negligible. Eddy-viscous damping in the internal zones is not present. NN and PP models calculated with $\alpha_m=0$ give qualitatively the same results. As eddy-viscous terms do not enter the static structure calculations, linear results are actually the same in both PP and NN models, as already mentioned. We have also checked that the presented discussion does not depend on the eddy viscosity form used in the model (see Section~5.2). If eddy-viscous pressure is used instead of $E_q$ and $U_q$ terms, qualitatively the same turbulent energy profiles are observed in PP models. Turbulent energies are generated through $-P_\nu\frac{dV}{dt}$ term which enters eq. (28) instead of $E_q$-term.
 
In the above discussion we considered the models with convective parameters of set A, that is without turbulent pressure, and what is more important without turbulent flux, which is, in the adopted model, responsible for the overshooting. Turbulent flux diffuses the turbulent energies beyond the convective instability regions, and hence significant turbulent energies may be present in convectively stable zones. This is physical overshooting. In the just  discussed models, turbulent flux was turned off. The NN models do not display significant overshooting, as turbulent energies are rapidly damped in convectively stable regions. Contrarily, PP models display high turbulent energies in convectively stable, internal zones, despite the fact that turbulent flux is turned off. This effect looks like overshooting, but in fact it is different from overshooting and we will call this phenomenon artificial overshooting. Turbulent energies are generated at the cost of pulsations (through the $E_q$-term) and are not effectively damped, because of the neglect of buoyant forces. Also, turbulent elements do not carry kinetic energy (turbulent flux turned off) nor heat, as convective flux is equal to 0 in convectively stable layers of PP model, by definition. Internal zones, with significant turbulent energies, leading to the additional eddy-viscous damping, cover no less than 6-7 local pressure scale heights (zones 40-70 in the discussed models). Thus, the range of artificial overshooting is extremely large. 

Test calculations done with turbulent flux turned on show, that all the conclusions from the above discussion remain unchanged. This is supported by Figs.~11 and 12. In Fig.~11 we show the Fourier decomposition parameters for models of set B. Solid line is for NN models, while dotted line for PP models. We observe the same differences as visible in Fig.~8 for models of set A. In Figs.~12 we plot the profiles of turbulent energy for 4.5\MS{} model of set B calculated with PP convection model. These are to be compared with Figs.~9c,d.  Distribution of turbulent energies in the deep interior (Fig.~9c~and~12a) is qualitatively the same for models with and without turbulent flux. It is still shaped by the balance of $D$ and $E_q$-terms. The range of this artificial overshooting may be larger than in case without turbulent flux, but it cannot be smaller than just estimated value of $\sim$(6-7) local pressure scale heights. Differences are visible in the external zones. Comparing Figs.~9d and 12b one sees smoother profiles of turbulent energies in model with turbulent flux, and higher turbulent energies in the outermost zones. These effects are caused by physical overshooting.

There are also strictly numerical consequences of different treatments of the source function in NN and PP convective models. During the integration of PP model, we deal with few orders of magnitude in $e_t$, only (Fig.~9d). Convergence is relatively fast, and constant time-step may be used thorough the whole model integration. Price to pay for the correct treatment of convectively stable regions in NN models, is greater numerical cost. NN models deal with turbulent energies spanning many orders of magnitude. If convective instability arises, turbulent energy must grow by several orders of magnitude, as is visible in Fig.~9b. This leads to convergence difficulties, and sometimes, the chosen time-step needs to be shortened, as discussed in Section~3.3.

\subsubsection{ $S\sim Y$ and $F_c\sim Y$ versus $S\sim Y$ and $F_c\sim Y_+$}

From the discussion presented in the proceeding Section, it is clear that it is the different treatment of the source function in convectively stable regions, that leads to the described differences in computed NN and PP models. Treatment of convective flux, that is  $F_c\sim Y$ versus $F_c\sim Y_+$ plays a minor role. This is easy to understand. PP and NN models differ in convectively stable regions. In PP model, convective flux is equal to zero in convectively stable zones by definition and in NN model, turbulent energies are very small there and thus, convective flux is negligible anyway. This is fully supported by the model sequence calculated with $S\sim Y$ and $F_c\sim Y_+$ (model NP) - dashed line in Figs.~8 and 11. In Fig.~8 NP sequence is almost overlapped with solid lines for NN sequence. Slightly higher difference between NN and NP models is visible in Fig.~11, where models with turbulent flux are displayed. Physical overshooting present in these models lead to higher turbulent energies in convectively stable zones, and hence effects of negative flux are stronger. However, the range of internal overshooting in NN and NP models is small, compared to PP models, and therefore, also the region with significant negative convective flux is very thin. We note that negative convective flux is always very small in our models, never exceeding 10 per cent of the total flux, usually being much smaller.

\subsubsection{Physical interpretation of negative $S$ and $F_c$}

It is clear from Kuhfu\ss{} (1986) derivation, that there is no need to restrict the turbulent source function, as well as convective flux to non-negative values. Below we show, that source function is proportional to forces acting on turbulent eddy during its motion. As these forces act both in convectively stable, and convectively unstable zones, there is no justification to neglect them in the former case.

Our reasoning follows the MLT considerations, concerning the acceleration of convective elements. For details of the derivations, we refer the reader to Cox \& Giuli (1968, \S 14.3). Below we assume that the convective eddy moves adiabatically, which corresponds to the neglect of radiative losses.

For the turbulent eddy, equation of motion may be written as follows
\begin{equation}\ddot{r}=-g-\frac{1}{\rho}\frac{\partial P}{\partial r}=-g\frac{\Delta \rho}{\rho},\end{equation}
where $\Delta \rho$ is excess density of the turbulent element. Thus, acceleration of the element, resulting from buoyant forces is equal to
\begin{equation}a=-g\frac{\Delta \rho}{\rho}.\end{equation}
Excess density may be expressed through the excess temperature, $\Delta T$, $(\Delta \rho/\rho)=-T\rho Q(\Delta T/T)$ which, after traveling the mixing length $\Lambda$, is
\begin{equation}\Delta T(\Lambda)=-\Lambda\frac{T}{P}\frac{dP}{dr}(\nabla-\nabla_\mathrm{a}).\end{equation}
Using above relations and equation of hydrostatic equilibrium to eliminate $g$ from eq. (30), we obtain a final equation for eddy acceleration
\begin{equation}a=\Lambda\frac{TQ}{P}\bigg(\frac{dP}{dr}\bigg)^2(\nabla-\nabla_\mathrm{a})=\alpha\frac{TPQ}{H_P}Y.\end{equation}
This is a buoyant acceleration of turbulent eddy displaced up by distance $\Lambda$. It is positive in convectively unstable region ($Y>0$) and negative in convectively stable region ($Y<0$). Comparison with definition of source function (14), leads to following relation
\begin{equation}S=\alpha_s e_t^{1/2}a\sim Y.\end{equation}
Thus, the source function is proportional to the acceleration of convective eddies caused by buoyant forces. This acceleration does not vanish in the convectively stable regions (it becomes negative). Therefore, there is no reason to set the source function to zero in convectively stable regions of the stellar model. To the contrary, source function has to be negative, because buoyant forces are now slowing down motion of the convective eddies. Assuming $S=0$ in convectively stable regions amounts to neglecting buoyancy, which is physically incorrect.

Overshooted elements carry kinetic energy (turbulent flux), as well as thermal energy (convective flux), which are dissipated in convectively stable zones. It is easy to show, that in convectively stable regions of the star, convective flux has to be negative, just as the source function. Indeed, when convective eddy overshoots down from the envelope convective zone, it becomes hotter than the surrounding medium (because $\nabla_\mathrm{a}>\nabla$). When eddy overshoots up, it becomes cooler than the surrounding medium. In both cases, convective flux is directed downwards ($F_c<0$).

For further support for negative source function and negative turbulent flux we refer the reader to Kuhfu\ss{} (1986), Gehmeyr \& Winkler (1992a,b), Canuto (1997). 

\section{Conclusions}

In this paper we describe, our convective hydrocodes for radial stellar pulsation. Convection model we use, is based on the Kuhfu\ss{} (1986) model. In the first part of the paper we briefly describe the model and list all model equations and necessary quantities. Technical details, concerning numerical schemes and methods we use, are fully described in Section~3 and in Appendices. Many tests we have done, part of which are briefly described in Section~4, prove that our code works properly. Computed models are numerically robust and reproduce well basic features observed in classical Cepheids, like Hertzsprung bump progression.

There are several other hydrocodes, that adopt similar convection model as our code. However, in different hydrocodes different treatments of some quantities, such as the turbulent source function, $S$, or eddy-viscous terms are used. Consequences of these differences were not fully studied up to date. In Section~5 we compare some of these treatments. Our most important finding, concerns the treatment of source function in convectively stable zones. In our hydrocode, we allow for negative source function in convectively stable zones, which reflects negative buoyancy, and is physically well motivated. However, in some other codes (\eg Florida-Budapest code), source function is restricted to non-negative values. This corresponds to the neglect of buoyant forces in convectively stable layers of the model. Similarly, convective flux is also restricted to non-negative values. We find that such approach has several serious drawbacks, that we list below:

({\it i}) Due to neglect of negative buoyancy effects (by assuming non-negative source function), significant turbulent energies are present in convectively stable layers of the model. When negative source function is absent in convectively stable layers, balance between eddy-viscous driving and turbulent dissipation sets the turbulent energies at relatively high level, $10^9-10^{10}$erg/g. We call this phenomenon artificial overshooting, as turbulent energies are generated by pulsations. Also, turbulent eddies do not transport heat into convectively stable layers, as convective flux is equal to 0.

({\it ii}) The range of this artificial overshooting is very large. Significant turbulent energies extend to more than 6 local pressure scale heights below the envelope convection zone.

({\it iii}) Significant turbulent energies in the deep, convectively stable parts of the model, lead to strong eddy-viscous damping, and consequently to lower pulsation amplitudes of the models in comparison to models computed with negative source function.

({\it iv}) Physical overshooting, due to turbulent flux, plays a minor role, and described, high turbulent energies are present in convectively stable layers, regardless if turbulent flux is included in the model or not.

In the next paper (Smolec \& Moskalik 2008) we will show further consequences of assuming non-negative source function. We show crucial role of this assumption in double-mode Cepheid models computed with the Florida-Budapest code (Koll\'ath \etal 2002). We will explain in detail the mechanism that leads to double-mode behaviour and we will show, that this mechanism does not work if source term is treated properly, that is, if we allow for negative values of the source function.

\Acknow{We are grateful to prof. Wojciech Dziembowski for fruitful discussions, specially concerning Sections~2~and~5. Alosza Pamyatnykh is acknowledged for the permission to use the opacity interpolating subroutines. Differences  between our formulation, and formulation used in Florida-Budapest code were clarified during the RS stay in Budapest. We wish to thank Robert Szab\'o for computation of several models with the Florida-Budapest hydrocode, used for comparison with our models. We are grateful to Robert Szab\'o and Zoltan Koll\'ath for fruitful discussions concerning the Florida-Budapest hydrocode. Hospitality of Konkoly Observatory staff is acknowledged. This work has been supported by the Polish MNiSW Grant No. 1 P03D 011 30.}

\section{APPENDIX A: Numerical representation of the model: static case}

 The model envelope is divided into $N$ mass zones, separated by the interfaces. In our notation both zones and interfaces are denoted by integers. Adjacent interfaces of zone $i$ have indices $i-1$ (bottom interface) and $i$ (upper interface), just as in the scheme below: \\

\renewcommand{\arraystretch}{1.}
\noindent\begin{tabular}{ccl}
--------------- & $i+1$ &\\
   $i+1$        &       &\\
--------------- & $i$   & $M,\ DM2,\ U,\ R,\ H_P,\ Y,\ L_r,\ L_c,\ L_t,\ U_q$\\
   $i$          &       & $DM,\ T,\ V,\ Q,\ c_P,\ \kappa,\ P,\ P_t,\ E_q,\ S,\ D,\ D_r$\\
--------------- & $i-1$ &\\
\end{tabular}\\

\noindent Outermost zone, as well as outermost interface, have index $N$. All quantities are defined either at the zones or at the interfaces, just as presented in the scheme above. For some quantities spatial averages, denoted by curly brackets, need to be calculated. Average of the zone quantity (\eg temperature, $T$) is defined at the interface and average of the interface quantity (\eg radius, $R$) is defined at the zone, like in examples below
$$\sa{T}_i=0.5(T_{i}+T_{i+1}),$$
$$\sa{R}_i=0.5(R_{i-1}+R_{i}).$$
We also need to calculate the spatial differences, which are numerical representation of derivatives. We denote them by $\Delta$. Spatial difference of zone quantity (\eg pressure, $P$) is defined at the interface, and spatial difference of the interface quantity (\eg luminosity, $L$) is defined at the zone, like in examples below:
\begin{equation}\Delta P_i = P_{i+1}-P_{i},\end{equation}
\begin{equation}\Delta L_i = L_{i}-L_{i-1}.\end{equation}

Model builder solves the static version of equations (1)-(4).  These are rewritten in a Lagrangean form. Instead of internal energy equation (2), we solve the total energy equation (4). In the static limit this equation reduces to total luminosity conservation condition, $L=\mathrm{const}$. Complete set of equations is as follows

\begin{equation}0=-4\pi R_i^2\frac{\Delta P_i+\Delta P_{t,i}}{DM2_i}-\frac{GM_i}{R_i^2},\end{equation}
\begin{equation}0=\frac{L_{r,i}}{L}+\frac{L_{c,i}}{L}+\frac{L_{t,i}}{L}-1,\end{equation}
\begin{equation}0=-\frac{\Delta L_{t,i}}{DM_i}+C_i.\end{equation}
Turbulent energy equation is defined at the zones, while luminosity conservation and momentum equations at the interfaces. Mass enclosed by radius $R_i$ is denoted by $M_i$, mass of the zone $i$ by $DM_i$ and mass associated with the interface $i$ by $DM2_i$. We have
\begin{equation}DM2_i=0.5(DM_i+DM_{i+1}).\end{equation}
For given temperature, $T_i$, and volume, $V_i$, EOS procedure calculates other thermodynamic quantities defined at the zones: pressure, $P_i$, and energy, $E_i$ (both containing gas and radiation contribution), specific heat at constant pressure, $c_{P,i}$, and thermal expansion coefficient, $Q_i$. Opacity procedure calculates the opacity, $\kappa_i$, also defined at the zone. Remaining zone quantities are turbulent pressure (eq.~(6))
\begin{equation}P_{t,i}=\frac{\alpha_pe_{t,i}}{V_i},\end{equation}
and the quantities entering the coupling term, $C_i$ (eqs.~(12)-(14))
\begin{equation}S_i=\frac{T_iP_iQ_i}{c_{P,i}}\sab{\frac{\Pi}{H_P}}_ie_{t,i}^{1/2},\end{equation}
\begin{equation}D_i=\frac{\alpha_d}{\alpha}\frac{e_{t,i}^{3/2}}{\sa{H_P}_i},\end{equation}
\begin{equation}D_{r,i}=\frac{4\sigma\gamma_r^2}{\alpha^2}\frac{T_i^3V_i^2}{c_{P,i}\kappa_i\sa{H_P^2}_i}e_{t,i}.\end{equation}

Other quantities are defined at the interfaces. Pressure scale height is
\begin{equation}H_{P,i}=\frac{R_i^2}{GM_i}\sa{PV}_i.\end{equation}
Numerical representation of the superadiabatic gradient (eq.~(16)) is based on the formula given by Stellingwerf (1982)
\begin{equation}Y_i=\frac{4\pi R_i^2}{DM2_i}\frac{H_{P,i}}{\sa{V}_i}\bigg(\sab{\frac{Q}{c_{P}}}_i(\Delta P_i)-(\log T_{i+1}-\log T_i)\bigg).\end{equation}
Numerical representation of the $\Pi$ term is
\begin{equation}\Pi_i=\alpha\alpha_s\sa{c_P}_iY_i,\end{equation}
or if the flux limiter is turned on
\begin{equation}\Pi_i=\mathcal{F}_iF_L\bigg[\frac{\mathcal{G}_i}{\mathcal{F}_i}\bigg],\end{equation}
where:
\begin{equation}\mathcal{F}_i=\sqrt{\frac{2}{3}}\sab{\frac{E+PV}{T}}_i,\end{equation}
\begin{equation}\mathcal{G}_i=\alpha\alpha_s\sa{c_P}_iY_i.\end{equation}
Note, that in comparison to equations (15) or (19) we dropped the term $e_t^{1/2}$, which now appears separately in the definitions of source term, $S_i$, and in the convective luminosity definition below, where it is averaged at the interface. Our motivation was to assure that all quantities entering the coupling term in zone $i$, $C_i$, depend on the turbulent energy in this zone only, $e_{t,i}$. For the luminosities we have
\begin{equation}L_{c,i}=4\pi R_i^2\frac{\alpha_c}{\alpha_s}\sab{\frac{T}{V}}_i\Pi_i\sa{e_t^{1/2}}_i,\end{equation}
\begin{equation}L_{t,i}=-\frac{2}{3}\alpha\alpha_t(4\pi R_i^2)^2H_{P,i}\sab{\frac{1}{V^2}}_i\frac{e_{t,i+1}^{3/2}-e_{t,i}^{3/2}}{DM2_i},\end{equation}
\begin{equation}L_{r,i}=-\frac{4\sigma}{3}\frac{(4\pi R_i^2)^2}{DM2_i}\frac{T_{i+1}^4/\kappa_{i+1}-T_i^4/\kappa_i}{1-\frac{\log(\kappa_{i+1}/\kappa_i)}{\log(T_{i+1}^4/T_i^4)}}.\end{equation}
Averaging scheme in the expression for radiative luminosity comes from Stellingwerf (1975).

Eddy-viscous terms are not present in the construction of static envelope, since they depend on $U$. However, they appear in the linear as well as nonlinear code. For completeness we present their numerical representation below. $U_q$ term is defined at the interface, while $E_q$ at the zone
\begin{equation}E_{q,i}=4\pi X_i\frac{U_i/R_i-U_{i-1}/R_{i-1}}{DM_i},\end{equation}
\begin{equation}U_{q,i}=\frac{4\pi}{R_i}\frac{X_{i+1}-X_i}{DM2_i},\end{equation}
where $X_i$ is zone quantity defined as follows
\begin{equation}X_i=\frac{16}{3}\pi\alpha\alpha_m\frac{1}{V_i^2}e_{t,i}^{1/2}\sa{H_P}_i\sa{R^6}_i\frac{U_i/R_i-U_{i-1}/R_{i-1}}{DM_i}.\end{equation}

As an option alternative to $U_q$ and $E_q$ terms, we implemented the eddy-viscous pressure, defined, like other pressure terms, at the zone
\begin{equation}P_{\nu,i}=-\frac{16}{3}\pi\alpha\alpha_me_{t,i}^{1/2}\frac{1}{V_i^2}\sa{R^3H_P}_i\frac{1}{DM_i}\bigg(\frac{U_i}{R_i}-\frac{U_{i-1}}{R_{i-1}}\bigg).\end{equation}

\section{APPENDIX B: Numerical representation of the model: nonlinear scheme}

Let $Z$ denote any physical quantity entering our model. In our notation $Z_i^{(n)}$ stands for the value of $Z_i$ at some particular moment of time, denoted by upper index $(n)$. After the time step $DT$, its value is $Z_i^{(n+1)}$. Time difference will be denoted by capital $D$, for example $DU_i=U_i^{(n+1)}-U_i^{(n)}$. We also need to calculate the average value of $Z_i$ during the time step, which we denote by $\ta{Z_i}$. Usually we set
\begin{equation}\ta{Z_i}=\xi Z_i^{(n+1)}+(1-\xi)Z_i^{(n)},\end{equation}
with $\xi=1$ corresponding to fully implicit treatment and $\xi=0$ corresponding to fully explicit treatment. For some quantities, more complicated time averages are necessary, as required by energy conservation (see below).

Finite difference form of equations (1), (4), (3) and (5) is then

\begin{equation}\frac{DU_i}{DT}+4\pi\ta{R_i^2}\frac{\Delta\big(\ta{P_i}+\ta{P_{t,i}}\big)}{DM2_i}+GM_i\tabb{\frac{1}{R_i^2}}-\ta{U_{q,i}}=0,\end{equation}

\begin{equation}D(E_i+\varpi_i^2)+\big(\ta{P_i}+\ta{P_{t,i}}\big)DV_i+\frac{DT}{DM_i}\Delta\big(\ta{L_{r,i}}+\ta{L_{c,i}}+\ta{L_{t,i}}\big)-DT\ta{E_{q,i}}=0,\end{equation}

\begin{equation}D\varpi_i^2+\ta{P_{t,i}}DV_i+\frac{DT}{DM_i}\Delta\ta{L_{t,i}}-DT\ta{E_{q,i}}-DT\ta{C_i}=0,\end{equation}

\begin{equation}R_i^{(n+1)}=R_i^{(n)}+DT\ta{U_i}.\end{equation}
 
Numerical scheme used in nonlinear calculations, must preserve the total energy during the integrations. Appropriate scheme for the radiative hydrocodes was proposed by Fraley (1968). We consider the total energy of the model envelope, equal to
\begin{equation}E_{TOT}=\sum_{i=1}^{N}\bigg[DM_i(E_i+\varpi_i^2)+0.5DM2_iU_i^2-\frac{GM_iDM2_i}{R_i}\bigg],\end{equation}
and its change during one time-step, $D E_{TOT}/DT$. Energy conservation requires $D E_{TOT}/DT=0$.  Neglecting the eddy-viscous terms, $E_q$ and $U_q$, convective equations may be reduced to the form exactly corresponding to purely radiative case, with the following substitutions: $\tilde{E}=E+\varpi^2$, for the total energy, $\tilde{P}=P+P_t$, for the total pressure and $\tilde{L}=L_r+L_c+L_t$, for the total luminosity. Thus, without eddy-viscous terms, following time-averagings are necessary in order to preserve the total energy (Fraley 1968)

\begin{equation}\ta{U_i}=\frac{1}{2} U_i^{(n+1)}+\frac{1}{2}U_i^{(n)},\end{equation}
\begin{equation}\ta{R_i^2}=\frac{1}{3}\Big(R_i^{2\ (n+1)}+R_{i}^{(n+1)}R_i^{(n)}+R_{i}^{2\ (n)}\Big),\end{equation}
\begin{equation}\tab{\frac{1}{R_i^2}}=\frac{1}{R_i^{(n)}R_i^{(n+1)}}.\end{equation}

The way we average pressures and luminosities has no effect on energy conservation. In our code we adopt the values used in radiative codes, that is we set $\theta=\theta_t=1/2$ and $w_r=w_c=w_t=2/3$ in the following
\begin{equation}\ta{P_i}=\theta P_i^{(n+1)}+(1-\theta)P_i^{(n)},\end{equation}
\begin{equation}\ta{P_{t,i}}=\theta_t P_{t,i}^{(n+1)}+(1-\theta_t)P_{t,i}^{(n)},\end{equation}
\begin{equation}\ta{L_{r,i}}=w_r L_{r,i}^{(n+1)}+(1-w_r)L_{r,i}^{(n)},\end{equation}
\begin{equation}\ta{L_{c,i}}=w_c L_{c,i}^{(n+1)}+(1-w_c)L_{c,i}^{(n)},\end{equation}
\begin{equation}\ta{L_{t,i}}=w_t L_{t,i}^{(n+1)}+(1-w_t)L_{t,i}^{(n)}.\end{equation}

Taking eddy-viscous terms into account and adopting averaging scheme as described above, one finds, that the total energy changes during one time-step, by
\begin{equation}
\frac{D E_{TOT}}{DT} =\sum_{i=1}^{N}\bigg[DM_i\ta{E_{q,i}}+0.5DM2_i\ta{U_{q,i}}2\ta{U_i}\bigg].
\end{equation}
Thus, total energy is conserved (the above sum vanishes), if following definitions for $\ta{U_q}$ and $\ta{E_q}$ are adopted
\begin{equation}\ta{U_{q,i}}=\frac{4\pi}{\ta{R_i}}\frac{\ta{X_{i+1}}-\ta{X_i}}{DM2_i},\end{equation}
\begin{equation}\ta{E_{q,i}}=4\pi \ta{X_i}\frac{\ta{U_i}/\ta{R_i}-\ta{U_{i-1}}/\ta{R_{i-1}}}{DM_i},\end{equation}
\begin{equation}\ta{X_i}=\theta_u X_i^{(n+1)}+(1-\theta_u)X_i^{(n)},\end{equation}
\begin{equation}\ta{R_i}=\beta R_i^{(n+1)}+(1-\beta)R_i^{(n)}.\end{equation}
$\theta_u$ and $\beta$ are not restricted by energy conservation. We set $\theta_u=1$ and $\beta=1/2$.

Turbulent energy equation (60) is decoupled from the energy conservation analysis. For the coupling term we write
\begin{equation}\ta{C_i}=\gamma C_i^{(n+1)}+(1-\gamma)C_i^{(n)},\end{equation}
and set $\gamma=1$. Values of $\theta_u$ and $\gamma$ were chosen experimentally to assure fast convergence of the nonlinear iterations.

\section{APPENDIX C:  Eddy-viscous work integrals}

In this Appendix we present the derivations of non-linear and linear eddy-viscous work integrals. We follow the derivation method presented in Unno \etal (1989). Following these authors we rewrite momentum equation (1) in the vector form. For eddy-viscous terms we use relations (7) and (8). Momentum and energy equations are
\begin{equation}\frac{d\vec{U}}{dt}=-\frac{1}{\rho}\nabla P_{\mathrm{tot}}-\nabla\phi+4\pi\frac{1}{R}\frac{\partial X}{\partial M}\hat{e_r},\end{equation}
\begin{equation}\frac{d(E+e_t)}{dt}+P_{\mathrm{tot}}\frac{dV}{dt}=-\frac{1}{\rho}\nabla \vec{F}+4\pi X\frac{\partial (U_R/R)}{\partial R},\end{equation}
where $\hat{e_r}$ is unit vector in the radial direction, $\phi$ is gravitational potential, $U_R$ is radial (only) component of the velocity vector $\vec{U}$. $P_{\mathrm{tot}}=P+P_t$ and $\vec{F}=\vec{F_r}+\vec{F_c}+\vec{F_t}$. We multiply the momentum equation by $\rho\vec{U}$ to obtain the equation of conservation of mechanical energy
\begin{equation}\rho\frac{d(\vec{U^2}/2)}{dt}=-\vec{U}\nabla P_{\mathrm{tot}}-\rho\vec{U}\nabla\phi+4\pi\rho\frac{U_R}{R}\frac{\partial X}{\partial M}.\end{equation}
Adding this equation to energy equation multiplied by $\rho$, and using continuity equation ($\rho (dV/dt)=\nabla\vec{U}$) we get
\begin{equation}\frac{d(E+e_t+\vec{U}^2/2)}{dt}=-\vec{U}\nabla\phi-\frac{1}{\rho}\nabla (\vec{F}+P_{\mathrm{tot}}\vec{U})+4\pi\frac{\partial (XU_R/R)}{\partial R}.\end{equation}
Integrating over the mass of the envelope we get
\begin{equation}\frac{dE_{\mathrm{tot}}}{dt}=-\int_{M_0}^M\frac{1}{\rho}\nabla(P_{\mathrm{tot}}\vec{U}+\vec{F})dM+4\pi\int_{M_0}^M\frac{\partial}{\partial M}\Big(X\frac{U_R}{R}\Big)dM,\end{equation}
where $E_{\mathrm{tot}}$ is equal to the total energy of the envelope
\begin{equation}E_{\mathrm{tot}}=\int_{M_0}^M(E+e_t+\vec{U}^2/2+\phi/2)dM.\end{equation}
Note that the last integral vanish as at the inner boundary $U_R=0$ and at the outer boundary $X=0$. The remaining integral is rewritten in the following form
\begin{equation}\frac{dE_{\mathrm{tot}}}{dt}=-\int_{M_0}^M\frac{1}{\rho}\nabla(\vec{F})dM-\int_{r=R_0}P_{\mathrm{tot}}\vec{U}d\vec{s},\end{equation}
where $R_0$ is radius of the star. The last surface integral vanish due to our outer boundary condition ($P_{\mathrm{tot}}=0$). Using energy equation again, we obtain
\begin{equation}\frac{dE_{\mathrm{tot}}}{dt}=\int_{M_0}^M\bigg[\frac{d(E+e_t)}{dt}+P_{\mathrm{tot}}\frac{dV}{dt}-4\pi X\frac{\partial (U_R/R)}{\partial R}\bigg]dM.\end{equation}
Nonlinear work integral is obtained through integrating over the whole pulsation cycle, which yields
\begin{equation}W=\oint_0^P dt\int_{M_0}^M dM\bigg[P_{\mathrm{tot}}\frac{dV}{dt}-4\pi X\frac{\partial (U_R/R)}{\partial R}\bigg]=\oint_0^P dt\int_{M_0}^M dM\bigg[P_{\mathrm{tot}}\frac{dV}{dt}-E_q\bigg].\end{equation}
In the linear approximation we have
\begin{equation}W=\oint_0^P dt\int_{M_0}^M dM\bigg[(\delta P_{\mathrm{tot}})\frac{d(\delta V)}{dt}-4\pi (\delta X)\frac{\partial (U_R/R)}{\partial R}\bigg].\end{equation}
Using relation, $U_R=d(\delta R)/dt$, and assuming $\delta z=\Re(\delta ze^{i\omega t})$ for the perturbed quantities, after laborious but straightforward algebra we arrive at
\begin{equation}W=-\pi\int_{M_0}^M \Im\big[(\delta P_{\mathrm{tot}})^*(\delta V)\big]dM+\pi\int_{M_0}^M \Im\bigg[(\delta X)^*\bigg(\frac{\delta V}{R^3}-\frac{3V}{R^3}\frac{\delta R}{R}\bigg)\bigg]dM.\end{equation}
The first integral correspond to ordinary pressure work integral, and the second term is eddy-viscous work integral, if Kuhfu\ss{} form of eddy viscosity is used. In case of Koll\'ath form, eddy viscosity has a form of ordinary pressure, and is simply included in $P_{\mathrm{tot}}$.

\newpage
\begin{figure}
\includegraphics[width=6.1cm]{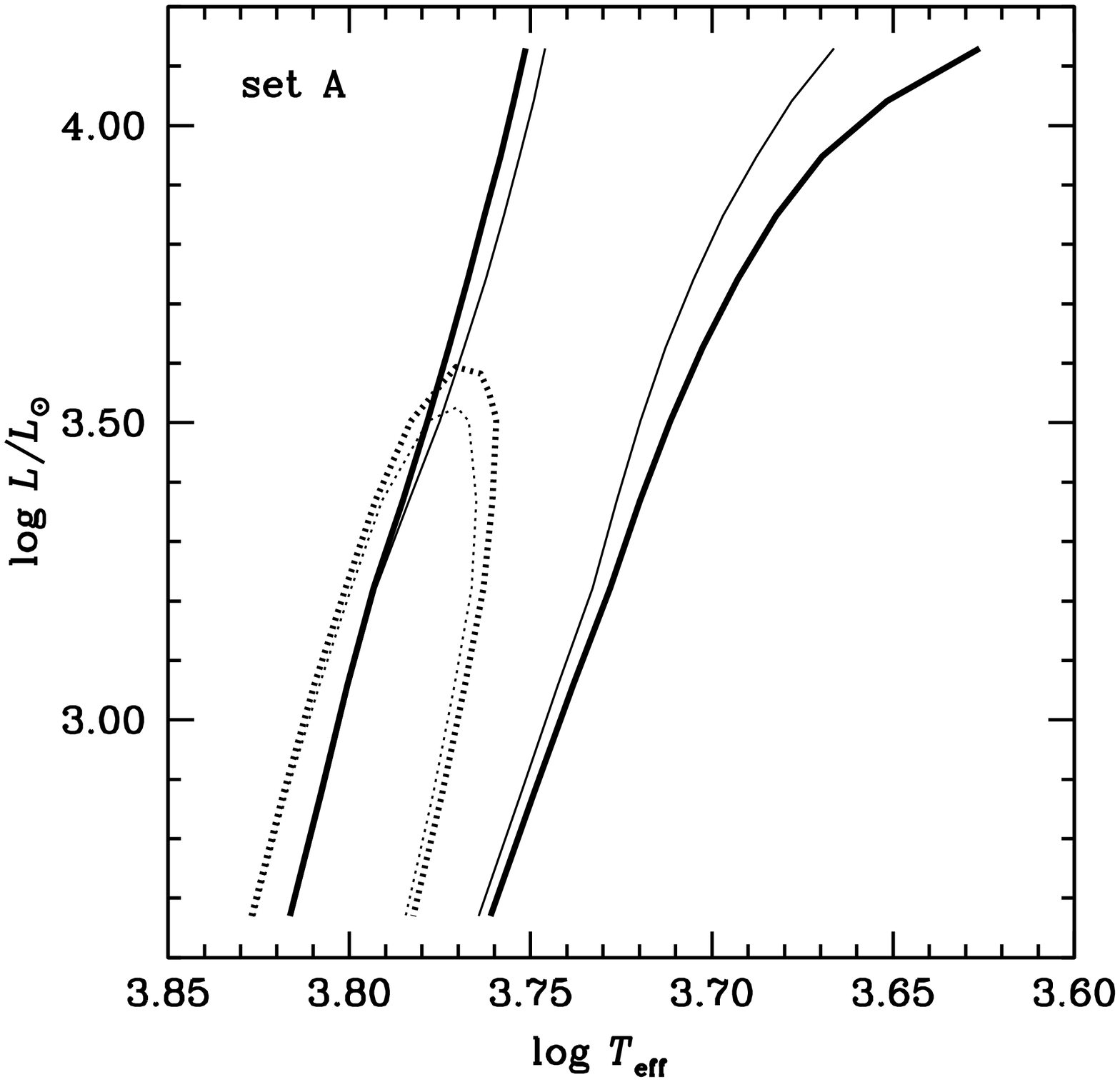}\includegraphics[width=6.1cm]{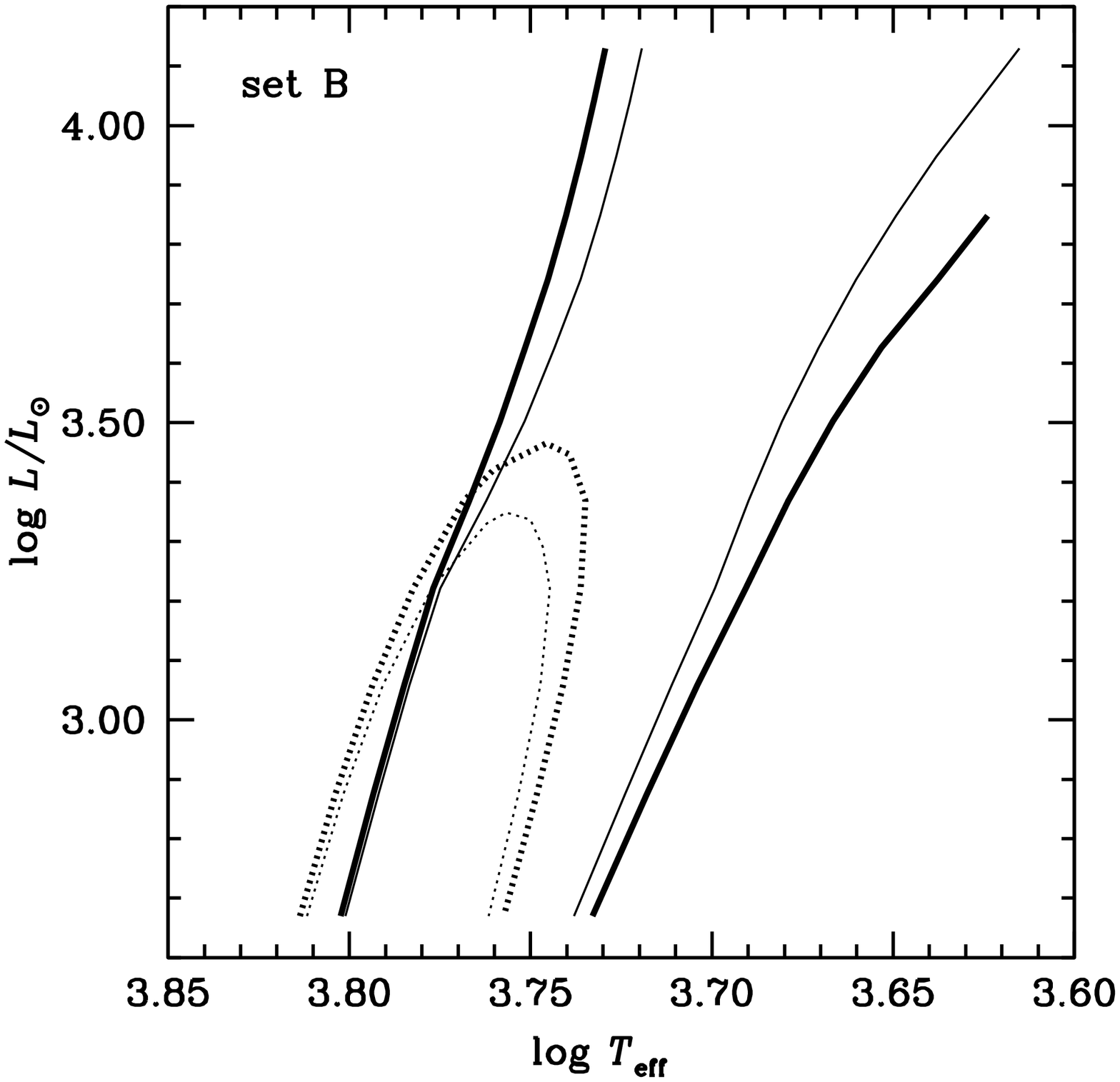}
\FigCap{Linear instability strips calculated for models with convective parameters of set A (left panel) and set B (right panel). Solid lines limit the fundamental mode instability strip, dotted lines enclose the first overtone instability strip. Thick lines refer to models calculated with our standard eddy viscosity form, while thin lines refer to models calculated with Koll\'ath eddy viscosity form, discussed in Section~5.2.}
\end{figure}

\begin{figure}
\includegraphics[width=11cm]{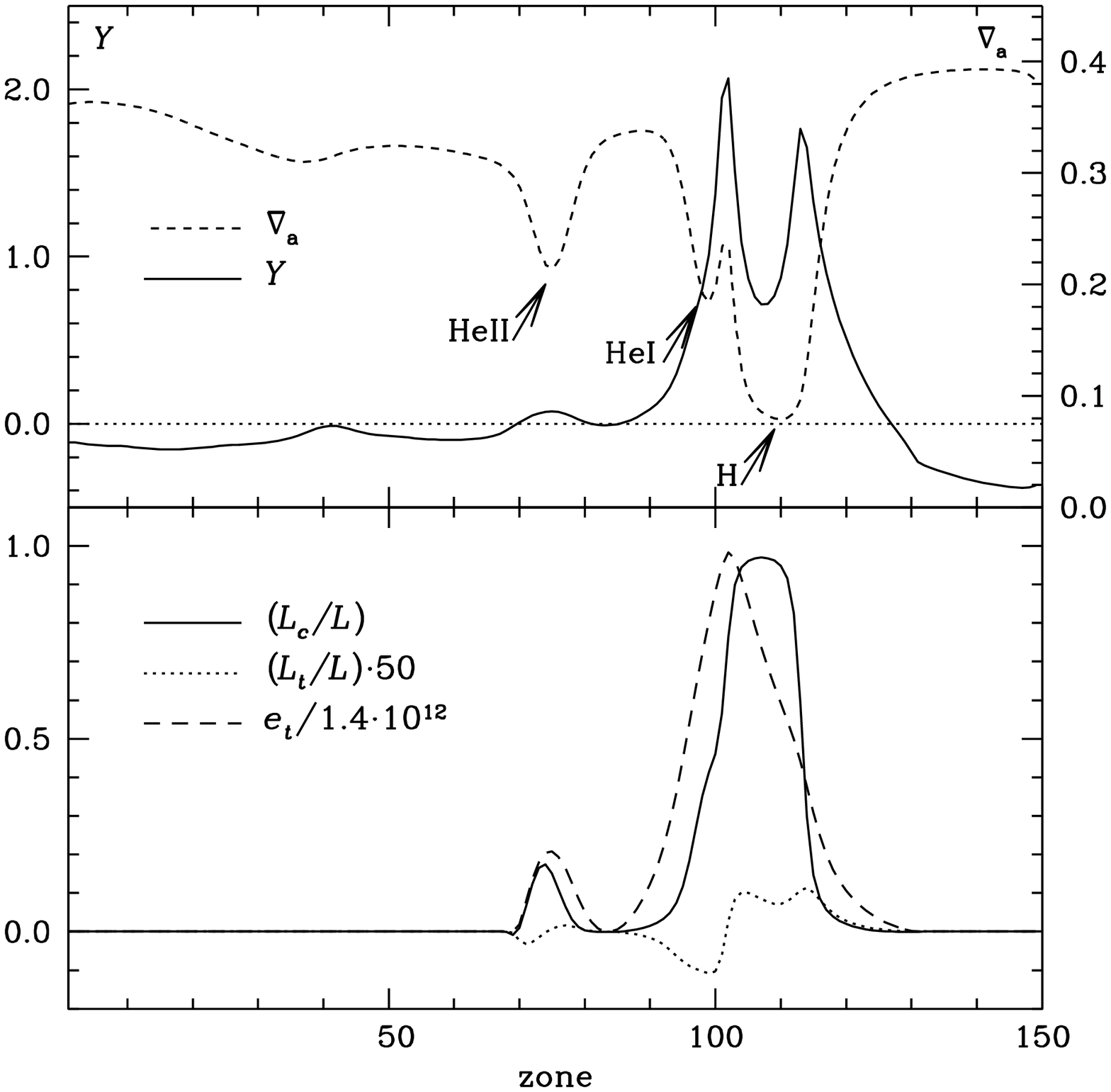}
\FigCap{Static structure for 4.5\MS{} model of set B, lying 300K to the red of the fundamental mode blue edge. All quantities are plotted versus the zone number. Surface at right. In the upper panel we display the run of $Y=\nabla-\nabla_\mathrm{a}$ and $\nabla_\mathrm{a}$. Labeled arrows mark the minima of $\nabla_\mathrm{a}$ connected with partial ionization regions of indicated element. In the lower panel relative convective and turbulent fluxes are plotted, together with scaled turbulent energy.}
\end{figure}

\begin{figure}
\includegraphics[width=11cm]{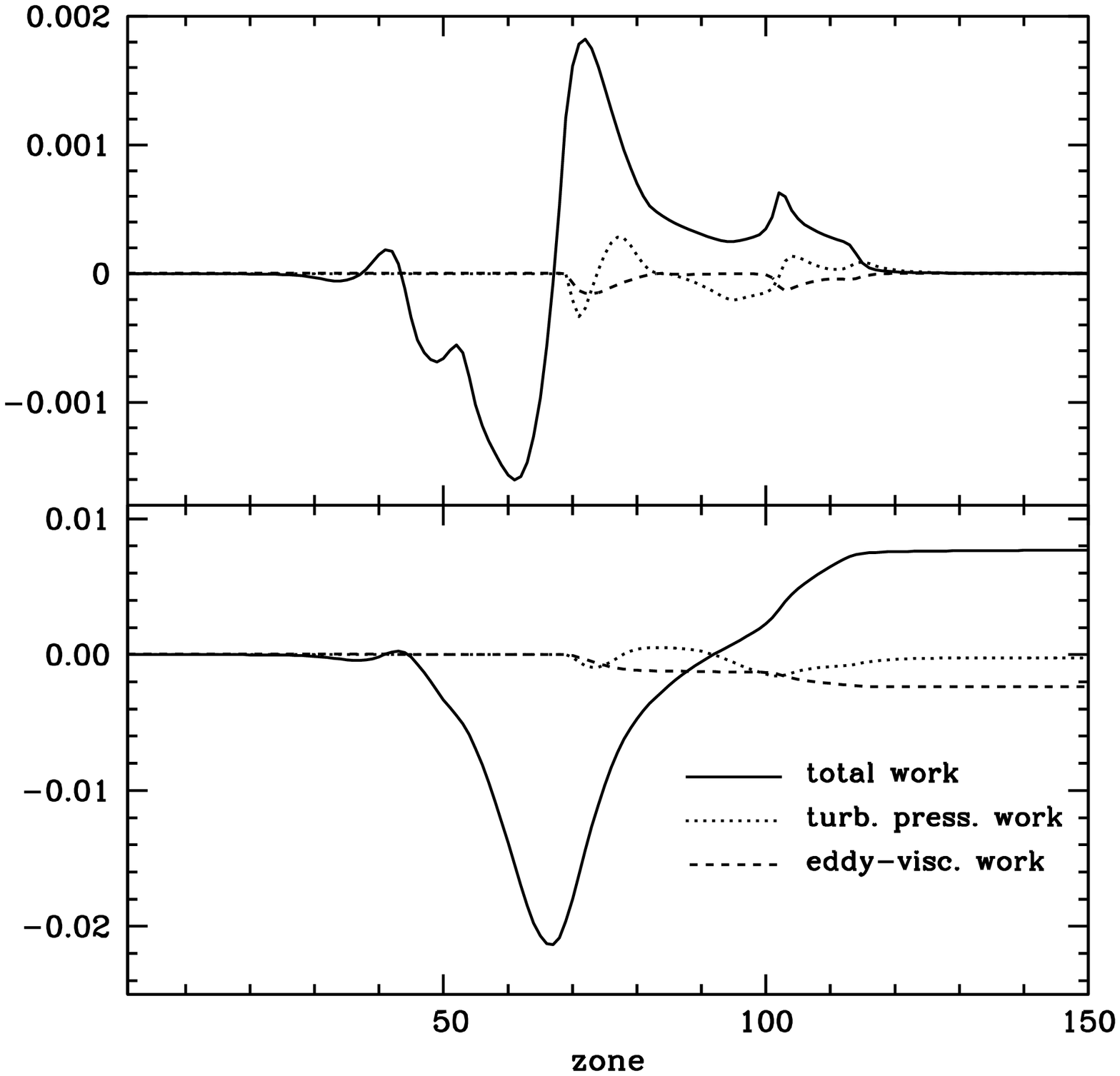}
\FigCap{Linear work integrals versus the zone number, for model of Fig.~2. Local work in the upper panel and cumulative work in the lower.}
\end{figure}

\begin{figure}
\includegraphics[width=11.cm]{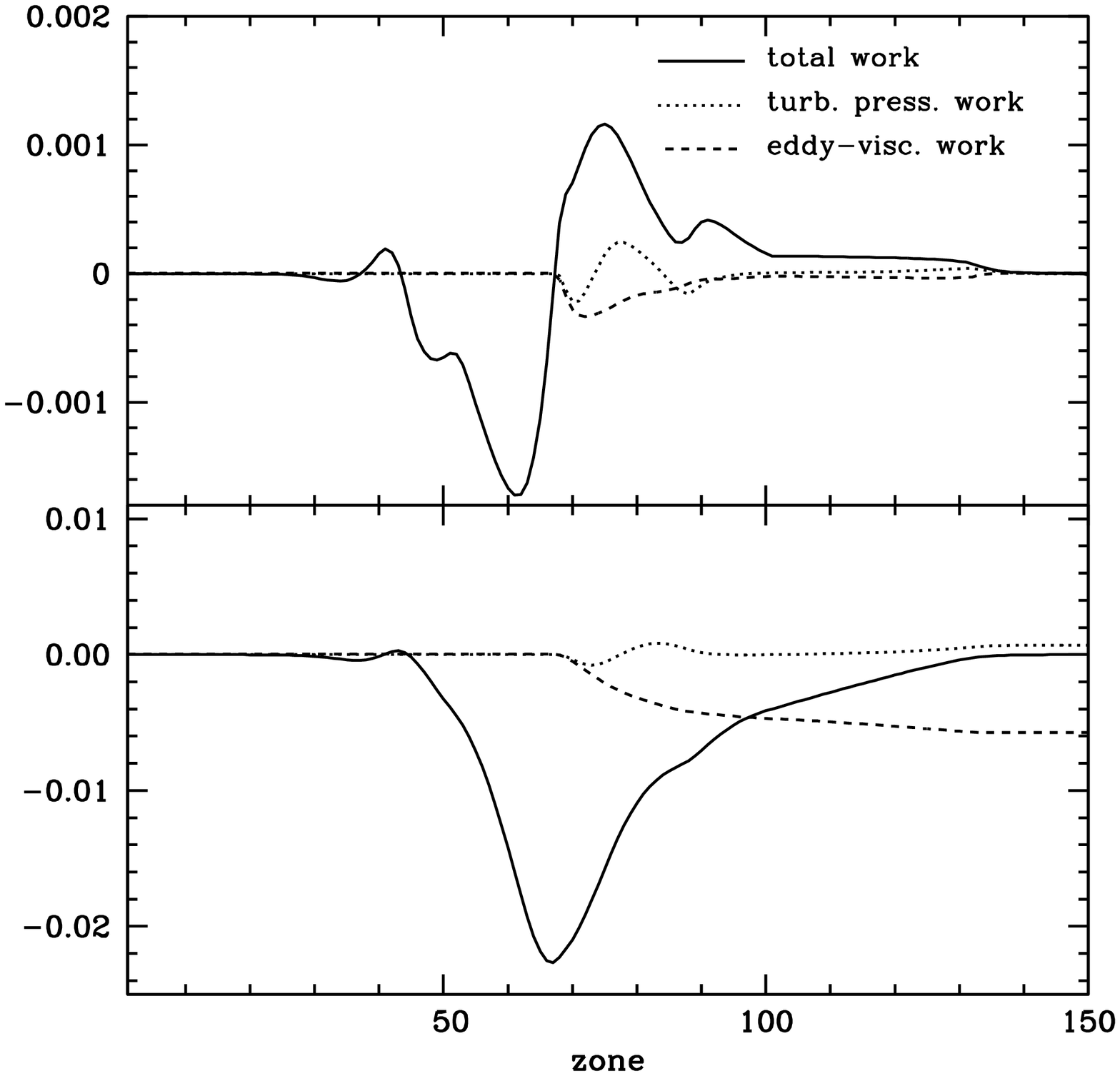}
\FigCap{Nonlinear work integrals versus the zone number, for model of Fig.~2. Local work in the upper panel and cumulative work in the lower.}
\end{figure}

\begin{figure}
\includegraphics[width=12.cm]{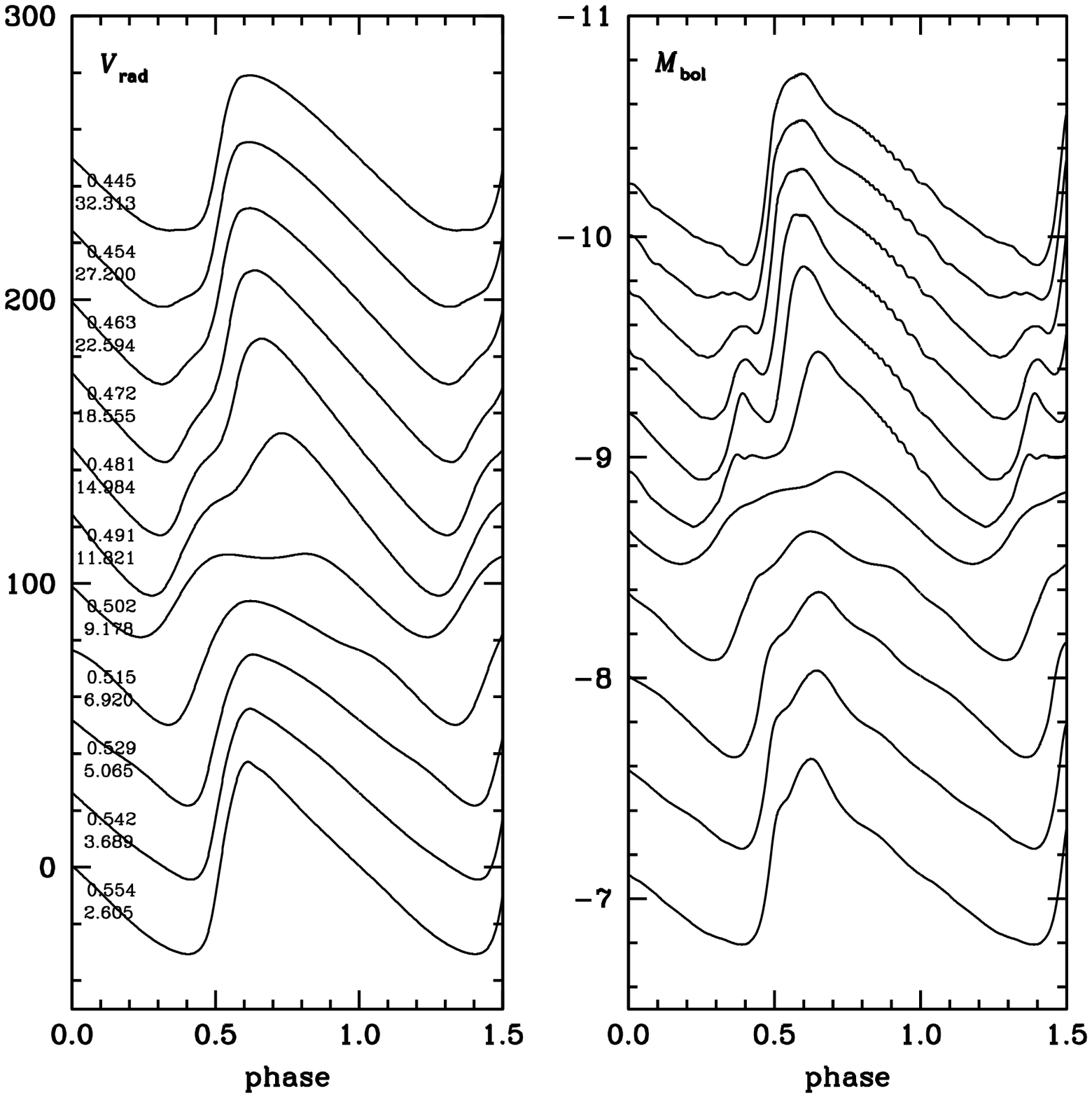}
\FigCap{Full amplitude radial velocity curves (left panel) and bolometric light curves (right panel) for models of set B, running parallel to the blue edge of the fundamental mode IS, 300K from it. Model masses are increasing by 0.5\MS, starting from 4.0\MS{} at the bottom of the figures up to 9\MS{} at the top. Consecutive radial velocity curves are shifted by 25 km/s to allow comparison. Curves are labeled by linear fundamental mode periods, $P_0$, and linear $P_2/P_0$ period ratios.}
\end{figure}

\begin{figure}
\includegraphics[width=12.cm]{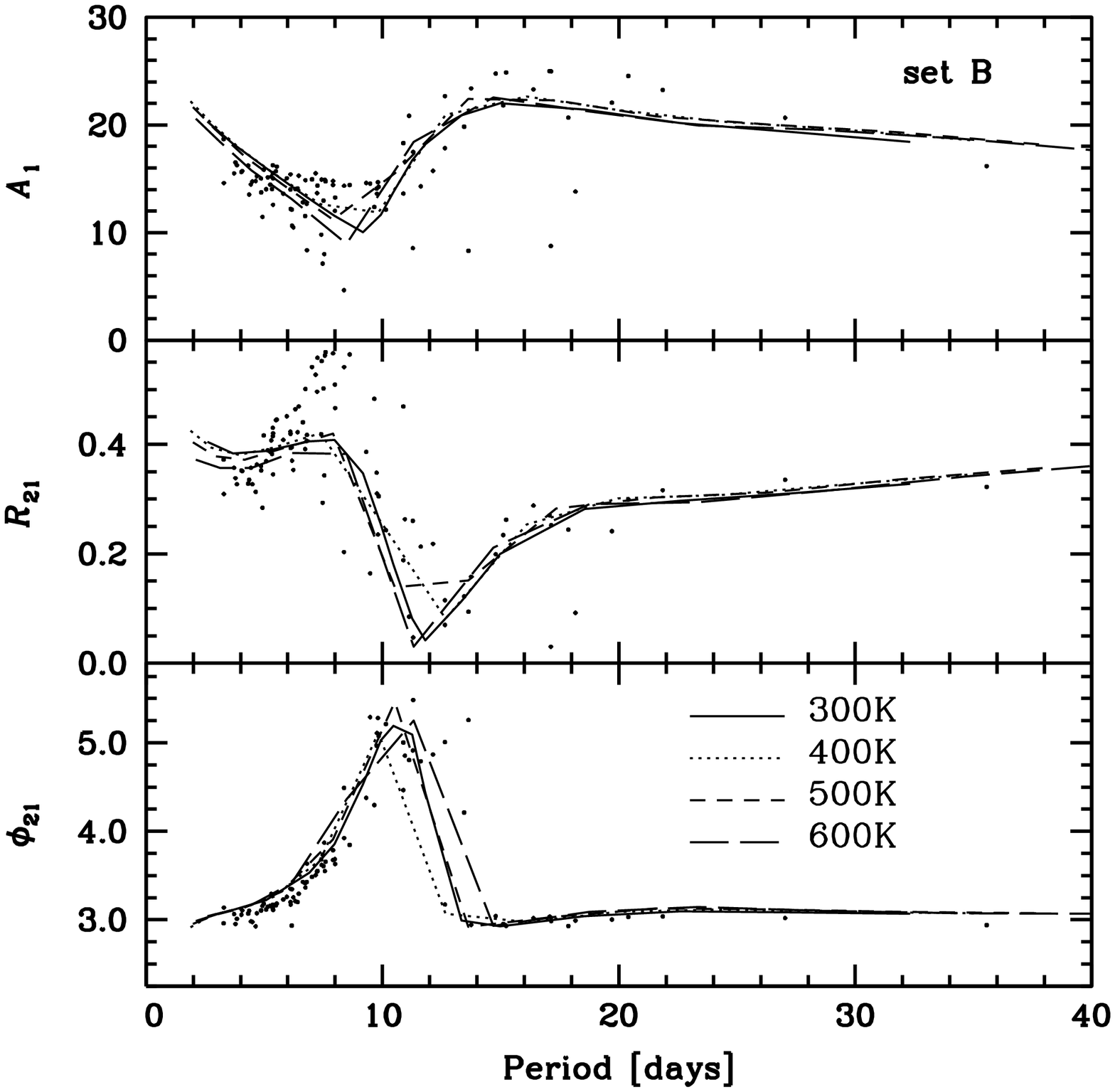}
\FigCap{Fourier decomposition parameters of radial velocity curves for models of set B, running parallel to the blue edge of the fundamental mode IS, 300K, 400K, 500K and 600K from it. Amplitudes are scaled by constant projection factor equal to 1.4. Individual curves for sequence running 300K apart from blue edge are presented in Fig.~5. Dots represent observational data (Moskalik, Gorynya \& Samus 2008, in preparation).}
\end{figure}

\begin{figure}
\includegraphics[width=12.cm]{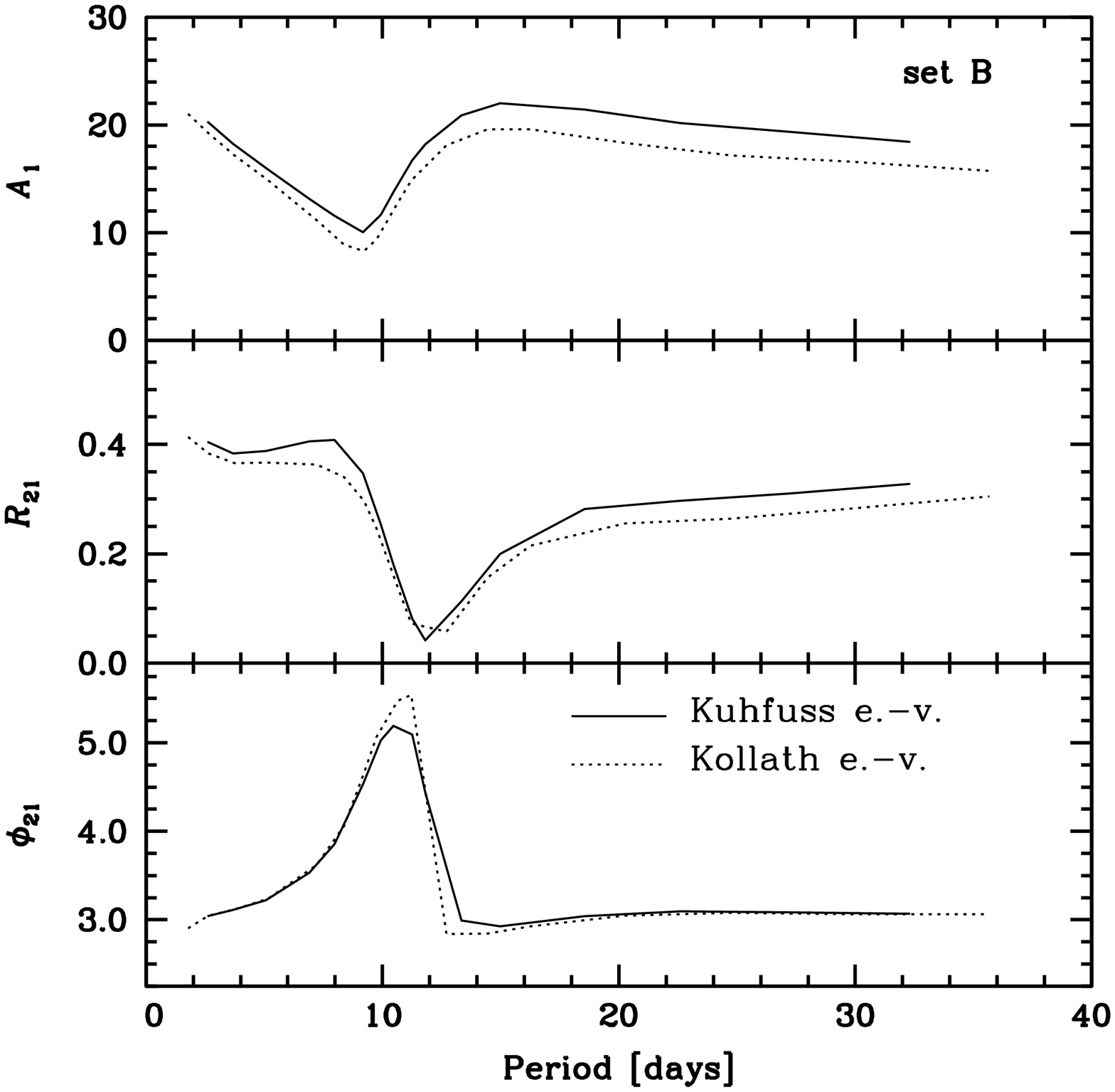}
\FigCap{Fourier decomposition parameters of radial velocity curves for models of set B, running parallel to the blue edge of the fundamental mode IS, 300K from it. Amplitudes are scaled by constant projection factor equal 1.4. Solid line for models with Kuhfu\ss{} eddy viscosity, dotted line for models with Koll\'ath eddy viscosity (see Section~5.2).}
\end{figure}

\begin{figure}
\includegraphics[width=12.cm]{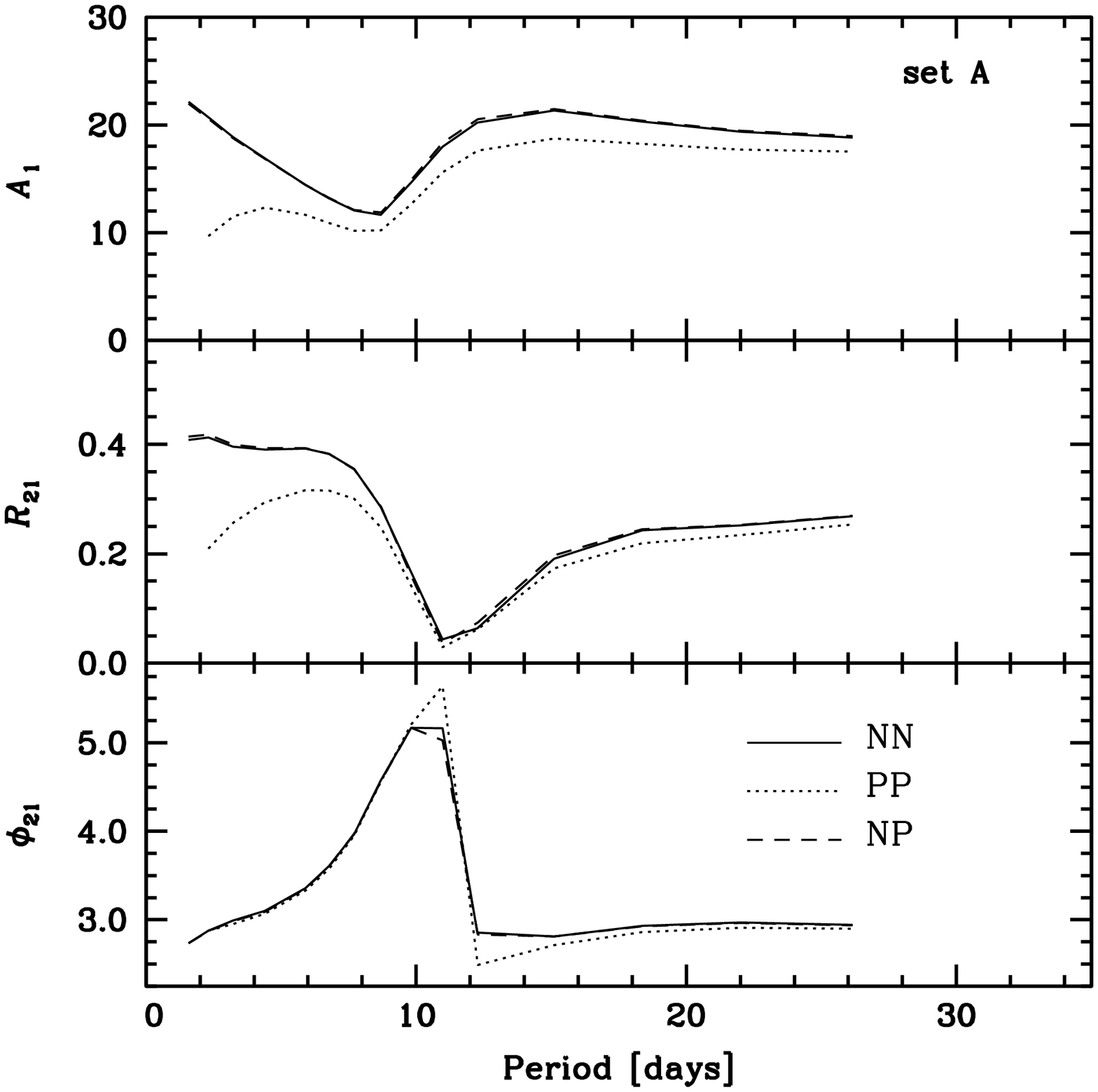}
\FigCap{Fourier decomposition parameters of radial velocity curves for models of set A, running parallel to the blue edge of the fundamental mode IS, 300K from it. Amplitudes are scaled by constant projection factor equal 1.4. Solid line for NN models, dotted line for PP models, and dashed line for NP models (see Section~5.3).}
\end{figure}

\begin{figure}
\includegraphics[width=12.2cm]{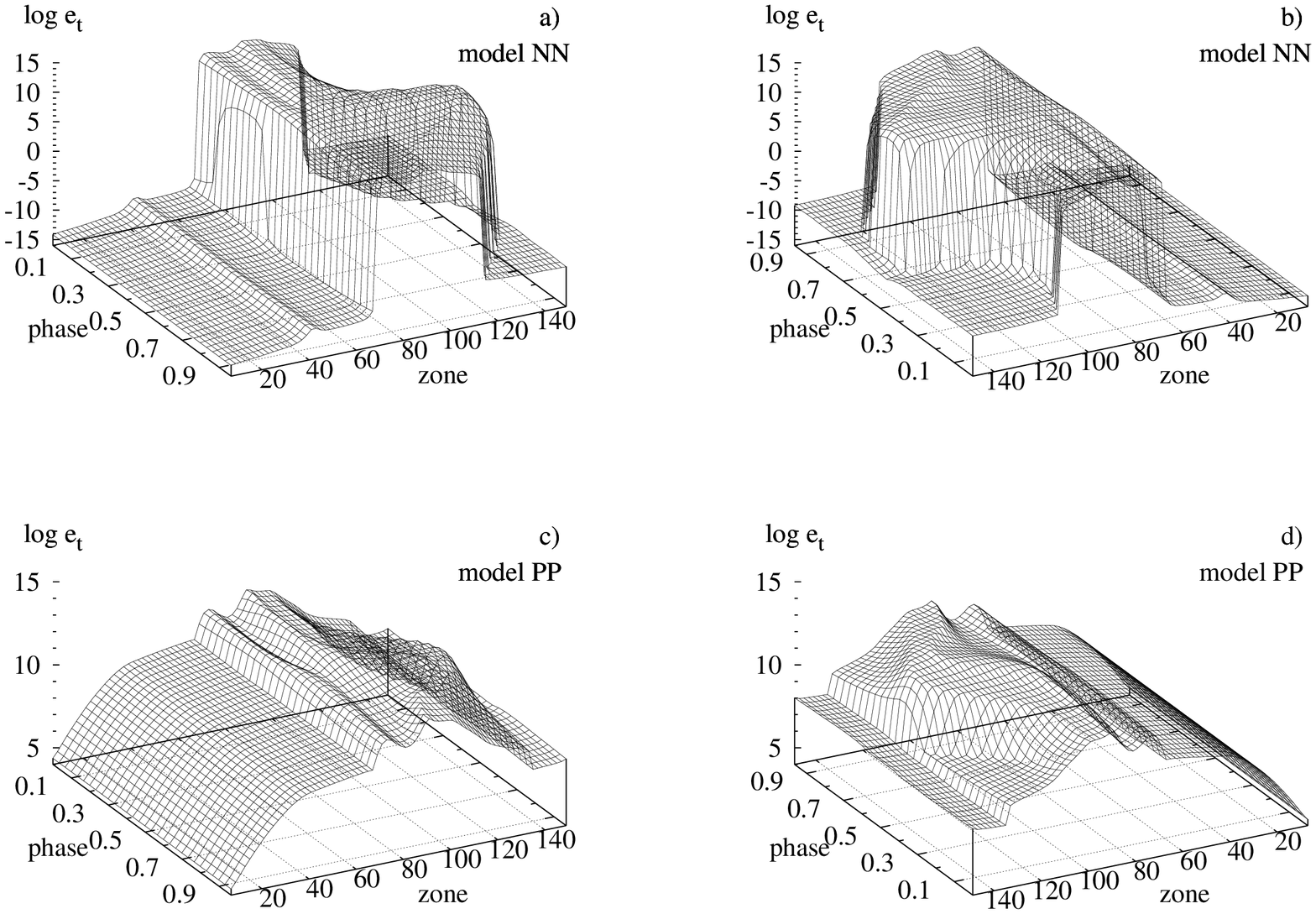}
\FigCap{Profiles of turbulent energy during one pulsation cycle for models discussed in Section~5.3. Panels a) and b) for NN model highlighting the internal and external parts of the model, respectively, while panels c) and d) present corresponding profiles for PP model.}
\end{figure}

\begin{figure}
\includegraphics[width=11.cm]{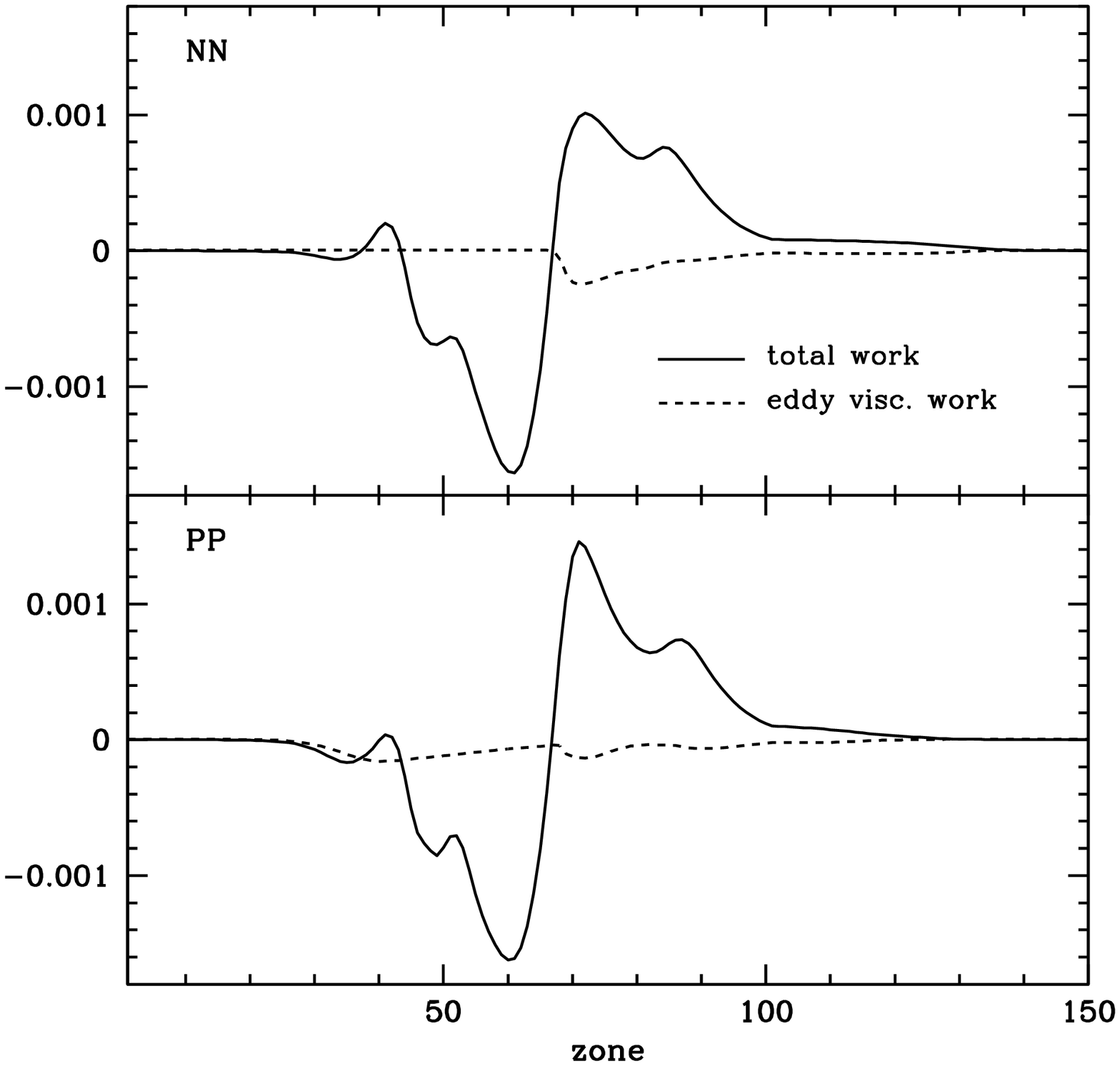}
\FigCap{Nonlinear, local work integrals versus the zone number, for 4.5\MS{} model of set A, lying 300K to the red of the fundamental mode blue edge. Upper panel for NN model, lower panel for PP model.}
\end{figure}

\begin{figure}
\includegraphics[width=12.cm]{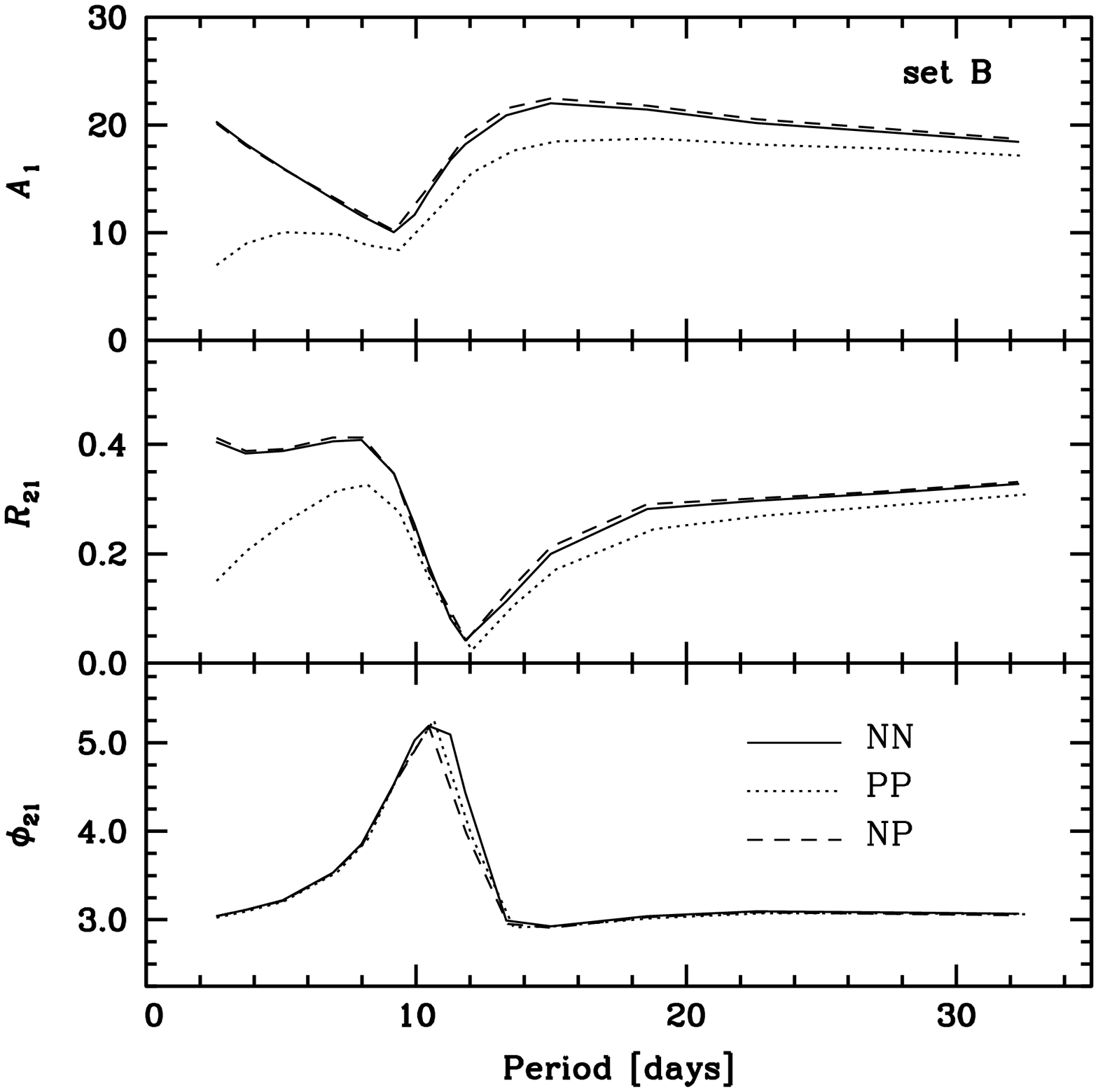}
\FigCap{Fourier decomposition parameters of radial velocity curves for models of set B, running parallel to the blue edge of the fundamental mode IS, 300K from it. Amplitudes are scaled by constant projection factor equal 1.4. Solid line for NN models, dotted line for PP models, and dashed line for NP models (see Section~5.3). }
\end{figure}

\begin{figure}
\includegraphics[width=12.2cm]{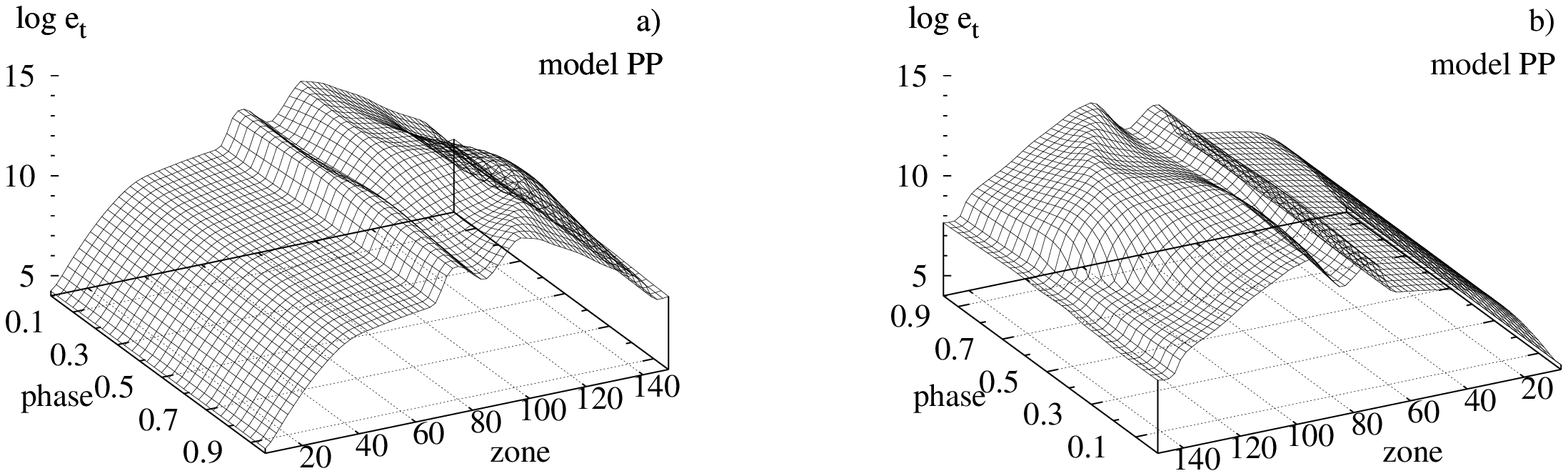}
\FigCap{Profiles of turbulent energy during one pulsation cycle for PP model with turbulent flux turned on, discussed in Section~5.3.}
\end{figure}

\end{document}